\documentclass[]{IEEEtran}
\IEEEoverridecommandlockouts

\usepackage{cite}
\usepackage{amsmath,amssymb,amsfonts}
\usepackage{algorithmic}
\usepackage{graphicx}
\usepackage{textcomp}
\usepackage{xcolor}
\usepackage[colorlinks, linkcolor=color, anchorcolor=blue, citecolor=color]{hyperref}
\usepackage{subfigure}

\DeclareMathOperator*{\Maximize}{maximize}

\newcommand{\bh}{\mathbf{h}}

\newcommand{\bv}{\mathbf{v}}

\newcommand{\bw}{\mathbf{w}}
\newcommand{\bx}{\mathbf{x}}
\newcommand{\bX}{\mathbf{X}}

\newcommand{\by}{\mathbf{y}}

\newcommand{\bq}{\mathbf{q}}

\newcommand{\bW}{\mathbf{W}}
\newcommand{\bY}{\mathbf{Y}}

\newcommand{\bn}{\mathbf{n}}

\renewcommand{\frac}{\dfrac}

\definecolor{myOrange}{rgb}{1,0.5,0}
\definecolor{myGreen}{rgb}{0,0.5,0}

\newcommand{\diagg}{\operatorname{diag}}

\newcommand{\maxi}{\operatorname{maximize}}
\newcommand{\subj}{\operatorname{subject~to}}
\def\BibTeX{{\rm B\kern-.05em{\sc i\kern-.025em b}\kern-.08em
    T\kern-.1667em\lower.7ex\hbox{E}\kern-.125emX}}
\begin{document}

\definecolor{color}{RGB}{128,0,0}
\newcommand{\re}[1]{{\color{blue}#1}}

\title{Role of Deep Learning in Wireless Communications 
}

\author{Wei Yu, \IEEEmembership{Fellow, IEEE}, Foad Sohrabi, \IEEEmembership{Member, IEEE}, and Tao Jiang, \IEEEmembership{Graduate Student Member, IEEE}
\thanks{Manuscript to appear in {\it IEEE BITS the Information Theory Magazine}. Wei Yu and Tao Jiang are with The Edward S.~Rogers Sr.\ Department of Electrical and Computer Engineering, University of Toronto, Canada. 
(e-mails: weiyu@ece.utoronto.ca, taoca.jiang@mail.utoronto.ca)
Foad Sohrabi is with Nokia Bell Labs, New Jersey, USA. 
(e-mail: foad.sohrabi@gmail.com)
This work is supported by the Natural Sciences and Engineering Research Council (NSERC) via the Canada Research Chairs program.} }

\maketitle

\begin{abstract}
Traditional communication system design has always been based on the paradigm of
first establishing a mathematical model of the communication channel, then
designing and optimizing the system according to the model.  The advent of
modern machine learning techniques, specifically deep neural networks, has
opened up opportunities for data-driven system design and optimization. This
article draws examples from the optimization of reconfigurable intelligent
surface, distributed channel estimation and feedback for multiuser beamforming,
and active sensing for millimeter wave (mmWave) initial alignment to illustrate
that a data-driven design that bypasses explicit channel modelling can often
discover excellent solutions to communication system design and optimization
problems that are otherwise computationally difficult to solve. We show that by
performing an end-to-end training of a deep neural network using a large number 
of channel samples, a machine learning based approach can potentially provide significant
system-level improvements as compared to the traditional model-based approach 
for solving optimization problems. The key to the successful applications of
machine learning techniques is in choosing the appropriate neural network
architecture to match the underlying problem structure.
\end{abstract}


\section{Introduction}
Modern machine learning techniques, specifically deep neural networks (DNNs), have
enabled tremendous progress for diverse applications, ranging from speech
recognition, natural language processing, image classification, to data
analytics and self-driving cars, and many more. In this article, we ask the
following question: Is there a role for machine learning in physical-layer
wireless communications system design? If so, where do opportunities lie, and 
where would the potential benefits come from? 

Fundamental to the phenomenal success of the machine learning techniques across
a wide range of applications is its apparent universal ability to approximate
any functional mapping from an input space to an output space, given sufficiently
complex neural network structure and enough training data \cite{hornik1989multilayer}. 
In fact, common characteristics of application domains where
machine learning has made the most impact, are that the inputs to these tasks
are high-dimensional complex data, whose structure needs to be explored, while
the outputs of these tasks can either be categorical (e.g., classification,
segmentation, sentiment analysis) or have complex structures themselves 
(e.g., machine translation, image labelling). The field of machine learning has developed
myriad techniques to enable automatic feature extraction and to explore the
structure of the problem in order to efficiently train a DNN to
map the input to the desired output. The machine learning paradigm essentially
solves optimization problems by pattern matching.  This is a vastly different
philosophy as compared to the traditional model-based information theoretical
approach to communication system design.

This article aims to illustrate that machine learning has an important role
to play even in the physical-layer wireless communications, which has traditionally
been dominated by model-based design and optimization approaches. This is so
for several reasons: 
\begin{itemize}
\item First, traditional wireless communication design methodologies typically
rely on the channel model, but models are inherently only an approximation to
the reality.  In applications where the models are complex and the channels are
difficult to estimate, a data-driven methodology that allows the system design to
bypass explicit channel estimation can potentially be a better approach.
\item Second, modern wireless communication
applications often involve optimization problems that are high dimensional,
nonconvex, and difficult to solve efficiently. By exploiting the availability of
training data, a machine learning approach may be able to learn the
solutions of the optimization problems directly. This can lead to a more
efficient way to explore the nonconvex optimization landscape than the
traditional model-based optimization approaches. 
\item Third, traditional
communication system designs are based on the principle of source-channel
separation and the optimal design of compression and channel codes. But when
the encoder and the decoder are block-length and/or complexity constrained, or
when the overall communication scenario involves multiple transmitters and
multiple receivers, the optimal design of practical encoder and decoder is highly
challenging. In this realm, there is the potential for discovering better
source and channel encoders and decoders using machine learning, as many of
these code design problems boil down to solving optimization problems over the
codebook structure for which data-driven methods may be able to identify
better solutions more efficiently.
\end{itemize}

The field of machine learning for communication system design has exploded 
in recent years \cite{o2017introduction,qin2019deep,8755300,eldar2022machine}. We mention some of the references here, e.g., in source and channel coding \cite{8723589,8242643,kim2020physical}, waveform design \cite{aoudia2022waveform}, signal detection \cite{ye2017power,farsad2018neural,9735332}, resource allocation \cite{8444648,8664604,lee2019graph,shen2020graph,9448070, 9783100} and channel estimation \cite{he2018deep,8272484}, etc. 
This article does not attempt to do justice in surveying the entire
literature and the recent progress on this topic. Instead, we focus on the
questions of why and how machine learning can benefit wireless communication
system design by presenting the following three specific examples. 

First, we consider communication scenarios in which a naive parameterization
of the channel would involve a large number of parameters, thus making channel
estimation a challenging task. Specifically, we show that in a wireless
communication system involving a reconfigurable intelligent surface (RIS), comprising
of a larger number of reflective elements, a machine learning approach that
directly optimizes the reflection coefficients without first estimating the
channel can significantly improve the overall performance  \cite{9427148}. 

Second, we consider a distributed source coding problem in the context of
channel estimation and feedback for a massive multiple-input
multiple-output (MIMO) system, and show that short block-length code design for
distributed data compression with system-level objective is feasible and can 
result in significant performance improvements over the 
single-user data compression codebook design \cite{9347820}. 

Third, we use an active sensing problem for millimeter wave (mmWave) initial
alignment to illustrate the role of machine learning in exploring the
optimization landscape in a complex sequential learning problem  \cite{9724252}. 
We show that selecting the right neural network architecture to match the
problem structure is crucial for its success.

\section{Information Theoretical Approach to Communication System Design}

Information theory has been the guiding principle in the development of
communication system design in the past seventy years. The driving philosophy
in information theory has always been reductionist---putting it in words of a
famous quote: {\it everything should be as simple as possible, but no simpler}.
A celebrated example of this philosophy is the additive white Gaussian noise
(AWGN) channel model, in which the choice of the Gaussian noise distribution is
justified both by a central limit theorem argument based on the assumption that
the overall noise is comprised of many independent small components and by the
fact that the Gaussian distribution is the worst-case noise distribution for 
the additive channel.  The AWGN model is cherished in the research community
and has played a central role in many historical developments in communication
theory (e.g., from time-domain equalization, to orthogonal frequency-division
multiplex, to multiuser detection), in coding theory (e.g., from maximum
likelihood decoding, to Viterbi algorithm, to Turbo, low-density parity-check,
and polar codes), and in multiuser information theory (e.g., from multiple-access,
to broadcast, and to interference channel models).

The wireless channels are however much more complicated than the AWGN channel
model. The wireless channel can be frequency selective; it is inherently
time-varying; it often involves multiple users and multiple antennas.
Historically, communication engineers have invested heavily in developing
models for various types of wireless channels. These models are often based on the
physics of electromagnetic wave propagation; many of these models are
statistical in nature; these channel models have played an important role in the design, analysis,
performance evaluation, and standardization of generations of wireless systems \cite{3GPP_channel_model}.

Channel modelling is important in wireless communication engineering
because most modern wireless systems 
operate under the framework of first estimating the channel, then feeding back the
estimated channel to the transmitter, and finally optimizing transmission and
reception strategies to maximize the mutual information between the input and
the output. 
In this article, we argue however that this model-then-optimize approach is not necessarily always the best approach.

\section{From Model-Based Optimization to Learning-Based Design}

\begin{figure*}[t]
\centering
\includegraphics[width=12cm]{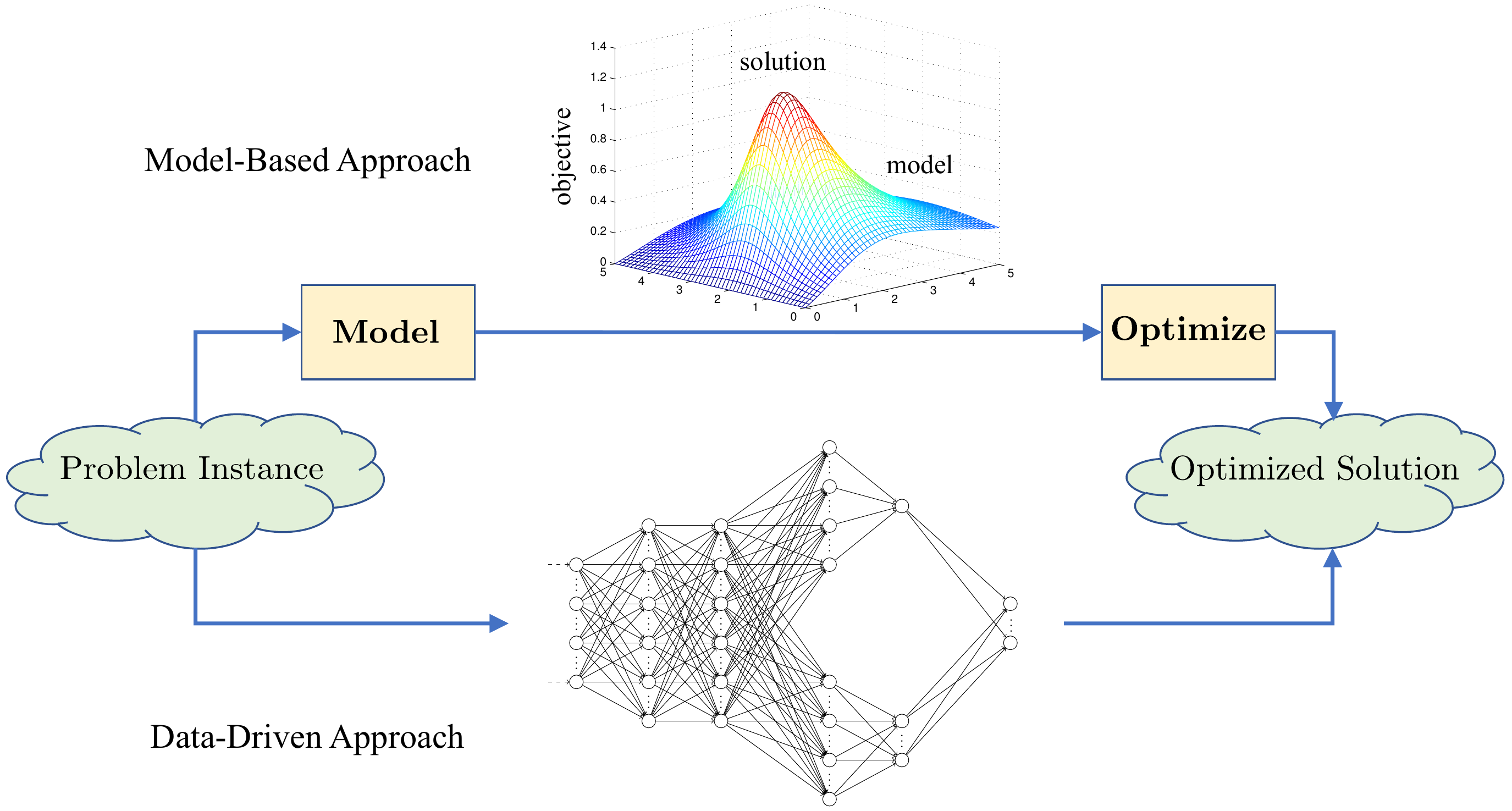}
\caption{Traditional wireless system design follows the paradigm of
model-then-optimize, as shown in the top branch. The design problem is modelled 
mathematically; the model parameters are then estimated, which allows the 
associated optimization landscape to be characterized; finally, the optimal
solution is obtained by mathematical programming. The machine learning approach
aims to directly learn the optimal solution from a representation of 
the problem instance, as shown in the bottom branch. The neural network is 
trained over many problem instances, by adjusting its weights according to 
the overall system objective as a function of the representation of 
the problem instances.}
\label{fig:ML_opt}
\end{figure*}

\subsection{Model-Based Communication System Design}

In traditional communication system design, maximizing the capacity of a wireless link typically requires channel estimation; the
process of channel estimation always depends on the channel model. Choosing which 
model to use is however an art rather than science.  This is because wireless channels
often have inherent structures that make certain models more appropriate than
others. For example, a MIMO channel with $M$
transmit antennas and $N$ receive antennas can simply be modelled as a $M\times
N$ matrix.  But a mmWave massive
MIMO channel often has a sparsity structure, corresponding to the finite number of propagation paths from the transmitter to the receiver, so that a
sparse path-based model in the spatial domain is a more efficient representation
of the channel.  Likewise, a frequency-selective channel can be modelled by its
channel response across the frequencies. But, the frequency selectivity is usually
a consequence of the different delays across the multiple paths, so the
channel variations across the frequencies are correlated. Instead of estimating
the channel in the frequency domain, a multipath time-domain model may be more
appropriate. 

Moreover, the channel estimation process requires specifying a loss function.
The squared-error metric is often adopted for tractability reasons, but
minimizing the mean-squared-error (MSE) of the estimates of the channel parameters
does not necessarily correspond to maximizing the overall system objective. 
For example, some parts of the channel may be more important
to describe than others. Clearly, the specific parameterization of the 
channel and the choice of the estimation error metric have a significant impact
on the ultimate system performance.

Traditionally, wireless researchers rely on experience and engineering
judgement in choosing the best channel model and the best optimization
formulation. The design decisions need to balance the inherent trade-offs
between: (i) how complex the model is, e.g., the number of parameters in the
model; (ii) how well the model approximates the reality; (iii) how easy it is to
estimate the model parameters; (iv) how easily the model can be used for
subsequent transmitter and receiver optimization.  We emphasize that in a
wireless fading channel with limited coherence time/frequency, model estimation
comes at a significant cost in term of the coherence slots occupied by pilot
transmissions. For example, a highly complex model may better approximate the
reality, but may require too many pilots for parameter estimation, hence may
not be worth the effort.  The point is that there is no universal theory about
how to choose the best channel model and how to best perform channel
estimation. To characterize and to take advantage of the underlying channel
structure in the design of the channel estimation process require
engineering intuition and are highly nontrivial tasks. 

In contrast, this article shows that a machine learning approach can
be used to allow an automatic discovery of the appropriate representation of
the channel based on training data. Further, it allows the optimization of the
system metric that actually matters (e.g., the achievable rate as opposed to
the MSE of the channel reconstruction) without having to first explicitly estimate the
channel.  This can have a
significant advantage as illustrated in the example of optimizing the RIS 
coefficients directly based on received pilots in Section \ref{sec:RIS} and 
the application of neural networks for 
channel feedback for the massive MIMO system in Section \ref{sec:CSI_feedback}. 


\subsection{Model-Based Optimization}

In many communication system design problem, even if the model parameters are
perfectly estimated, the resulting transmitter and receiver optimization
problem may still be not so easy to solve. 
The formulation of the optimization problem is also an art rather than science. 
In fact, wireless engineers often adopt optimization formulations, \emph{because}
the resulting mathematical programming problem is amendable to either analytic
or computationally efficient numerical solution. We remark that a
mathematical optimization problem can often be parameterized in many different
ways.  The ``holy grail'' of mathematical optimization is often thought of
as to transform a problem into a convex form, so that computationally efficient
numerical procedures can be developed to find the global optimal solution of
the resulting mathematical programming problem. But there is no universal
theory about how best to transform the optimization landscape. 

In contrast, this article shows that a machine learning approach can 
be used for the automatic discovery of the mapping from the problem 
representation to the optimal solution based on training data, as illustrated 
in the examples of optimizing RIS coefficients based on received pilots in Section
\ref{sec:RIS}, and optimizing of beamformers based on channel feedback
in Section \ref{sec:CSI_feedback}, finally optimizing a sequence of
active sensing strategies in Section \ref{sec:active_sensing}.

\subsection{Data-Driven Communication System Design}

The article advocates the viewpoint that a data-driven approach can
circumvent many of the modelling and optimization difficulties for wireless 
system design as mentioned in the previous section. The main idea is as shown 
in Fig.~\ref{fig:ML_opt}.  Instead of the traditional 
model-then-optimize approach, which involves choosing an appropriate parameter
space, then characterizing the associated optimization landscape, and finally
performing the resulting mathematical optimization, we adopt a data-driven
approach to directly map the problem instances to the corresponding optimized solutions. 
By training such a neural network over many problem instances, the task of
optimization is essentially turned into \emph{pattern matching}. When a new
optimization task comes along, the trained neural network can then simply output the
corresponding solution. This is akin to a human learner who is trained to use
past experience to perform future optimization tasks. 

The advantages of the proposed data-driven paradigm are:
\begin{itemize}
\item It allows direct system-level optimization without the intermediary
step of channel estimation.  The modelling uncertainty and the channel
estimation error are implicitly taken into account in the overall 
optimization process.  
\item It allows an end-to-end design with a realistic system-level objective
function, instead of relying on some arbitrary  metric in the model parameter estimation process. 
\item It allows the problem instances to be represented in an arbitrary
fashion. Additional side information which is often not easy to 
incorporate into a model can now be accounted for in the optimization process.
\item By using a large number of problem instances as training data, it allows 
the optimization process to efficiently explore the high-dimensional optimization 
landscape in the training stage. 
\item Once trained, the neural network can efficiently output the optimized
solution for new problem instances. In effect, the computational
complexity is moved from the optimization stage to the neural network training process. 
\end{itemize}

Thus, instead of using a mathematical optimization approach that requires
highly structured models over well-defined problems and relies on the specific 
(e.g., convex) structure of the optimization landscape, a machine
learning approach is capable of solving relatively poorly defined problems and
exploring high-dimensional optimization space without first identifying the
problem structure. This is made possible because of the ability of the neural
network to find patterns in the vast amount of training data, thanks to the nowadays
prevalent highly parallel computer architectures for both neural
network training and implementation processes \cite{GPU_2010,tensorflow2016}. 

Machine learning is about approximating functions---its broad impact comes from
the fact that it is particularly effective in processing high-dimensional data. 
The phenomenal success of deep learning in domains such as image and speech
processing is due to the fact that the specific task at hand is often governed by some low-dimensional characteristics 
(e.g., labels) embedded in high-dimensional observations (e.g., images). 
As we shall see in the examples in the sequel, the wireless communications
scenarios in which the data-driven optimization can be shown to substantially 
outperform the traditional model-driven design are also precisely the situations in which 
the problem instances have some low-dimensional structure and are observable 
only through limited number of high-dimensional outputs. In the communications setting, 
the observations are typically the received pilots; the low-dimension problem 
structure is typically due to the sparsity of the underlying wireless channel. 
The benefit of machine learning comes from bypassing the explicit modelling of
the channel structure and instead using a DNN to directly process the 
high-dimensional received pilots to arrive at a desired communication action. 
The remaining of this article uses three examples to illustrate the success of
machine learning in wireless applications. 




\section{Capacity Maximization for Reconfigurable Intelligent Surface System}
\label{sec:RIS}

\begin{figure*}[t]
    \centering
    \includegraphics[width=0.89\linewidth]{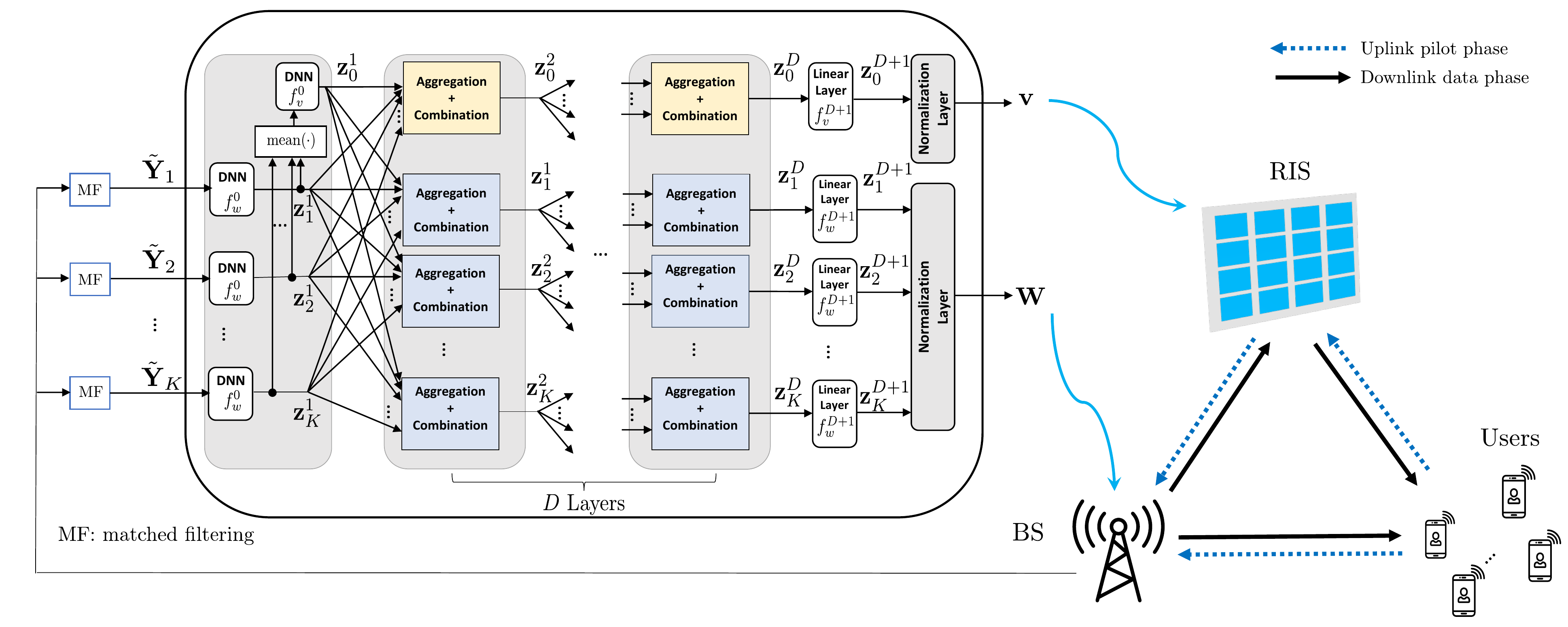}
	\caption{The deep learning framework for directly designing the multiuser beamformers and reflection coefficients based on the received uplink pilots for a downlink RIS-assisted multiuser system.}
    \label{fig_ris:dnn_overall_arch}
\end{figure*}

Wireless channels are often high dimensional. This is the case for massive
MIMO systems in which the transmitters and the receivers are equipped with large
antenna arrays, and is also true of emerging devices such as a class of metasurfaces
known as RIS, which consists of a large number of reflecting elements and can be
dynamically reconfigured to refocus the electromagnetic waves to the intended receivers \cite{di2020smart}. 

The physical electromagnetic propagation environment of a wireless channel
is also often sparse, especially as compared to the number of elements in
the antenna array or the reflective surface. This is because
the propagation characteristics typically only depend on a small number of scatters,
and the number of propagation paths in the environment can be significantly less
than the number of transmit, receive, or reflecting elements.
On the other hand, due to the limited number of radio-frequency (RF) chains and the finite
pilot overhead, the available observations of the channel is typically limited.

How can we estimate a sparse high-dimensional channel through limited number of 
observations?  The traditional approach is to take advantage of the channel
sparsity and to build a channel model with a small number of parameters, then
proceed with estimating the parameters of the channel based on the received
pilots, followed by optimizing the system according to the estimated channel.  
How well such an approach works would depend on how well the model approximates
the actual channel. In this section, we advocate an alternative data-driven approach
that bypasses the explicit modelling stage and directly optimizes the system
using a neural network with the received pilots as inputs. We use the RIS as
an example in which explicit channel estimation is especially challenging, but 
the proposed approach can be adopted equally well in many other scenarios, including the 
conventional massive MIMO system.



A commonly used model for RIS is to regard it as a device consisting of a large
number of tunable elements that can reflect incoming signals with arbitrary
phase shifts.  The goal is to dynamically reconfigure the phase shifts at the
RIS according to the channel realizations of the users in order to maximize a
system-level metric, e.g., the system downlink throughput. The problem is
that channel estimation is highly nontrivial for the RIS system. Assuming a
time-division duplex (TDD) system with channel reciprocity, channel estimation
can be done using uplink pilots. However, the large number of RIS
elements gives rise to a high-dimensional channel, which would require many
pilots to estimate. Further, even if the channel can be accurately estimated,
the optimization of the RIS coefficients is a highly complex and nonconvex
optimization problem, which is difficult to solve.

We show that the approach of using machine learning to directly map the
received pilots to the optimized RIS reflective coefficients can yield a
significant performance improvement as compared to the traditional channel
estimation based approach \cite{9427148}. The performance gain comes from the
fact that channel models are only an approximation of the reality and
that traditional channel estimation always needs to assume an estimation error
metric (such as the MSE), but such metric does not perfectly match
the system-level objective. This problem can be alleviated by bypassing the
modelling stage, by using the true system objective as the loss function, and
by training a neural network to directly output the optimized reflective
coefficients based on the received pilots. Essentially, the wireless channel 
is now represented by the received pilots. The complexity of high-dimensional 
optimization is shifted to the training stage, where a large number of 
channel instances and the corresponding reflecting coefficients are processed
by the neural network so that it can produce a desired solution when a new
channel realization is observed. 

Choosing the right architecture for the neural network turns out to be important. 
For this application, we experimentally find that the best system-level performance is obtained
by adopting a graph neural network (GNN) \cite{xu2019powerful,lee2019graph,shen2020graph} that captures the spatial relationship
between the {base-station (BS)}, the RIS, and the users. The proposed approach
and the interpretations of the solutions are presented below.


\subsection{System Model and Problem Formulation}
Consider an RIS-assisted {MIMO system} with a BS equipped with $M$ antennas serving $K$ single-antenna users. An RIS consisting of $N$ elements is deployed between the BS and the users to enable a reflection link. Let $\mathbf h_k^{\rm d}\in\mathbb{C}^{M}$ denote the direct channel from the BS to {user $k$}, and $\mathbf h_k^{\rm r}\in\mathbb{C}^{N}$ denote the channel from the RIS to user $k$, and $\mathbf G\in\mathbb{C}^{M\times N}$ denote the channel from the RIS to the BS. We assume a block-fading channel model.
In the downlink, the BS sends the data symbol $s_k\in\mathbb{C}$ with $\mathbb{E}[|s_k|^2]=1$ to {user $k$} using a {beamforming} vector $\mathbf w_k\in\mathbb{C}^M$, which satisfies a total power constraint $\sum_{k=1}^K\mathbb\|\mathbf w_k\|_2^2\le P_d$. The RIS reflection coefficients are denoted by $\mathbf v=[e^{j\omega_1},e^{j\omega_2},\cdots,e^{j\omega_N}]^\top$, where $\omega_i\in(-\pi,\pi]$ is the phase shift of the $i$-th element. Then, the received signal at user $k$ is represented as:
\begin{align}
    r_k &= \sum_{j=1}^K(\mathbf h_k^{\rm d}+ \mathbf G\diagg(\mathbf v)\mathbf h_k^{\rm r})^\top\mathbf w_j s_j + n_k \nonumber\\
        &= \sum_{j=1}^K(\mathbf h_k^{\rm d}+ \mathbf A_k\mathbf v )^\top\mathbf w_j s_j + n_k,
\end{align} 
where $\mathbf A_k = \mathbf G\diagg(\mathbf h_k^{\rm r}) \in\mathbb{C}^{M\times N}$ denotes the cascaded channel from the BS to user $k$ through reflection at the RIS, and $ n_k\sim\mathcal{CN}( 0,\sigma_0^2)$ is the additive white Gaussian noise.
The $k$-th user's achievable rate $R_k$ is computed as:
\begin{equation}
    R_k = \log_2\left(1+\frac{|(\mathbf h_k^{\rm d}+ \mathbf A_k\mathbf v )^\top\mathbf w_k|^2}{\sum_{i\neq k} |(\mathbf h_k^{\rm d}+ \mathbf A_k\mathbf v )^\top\mathbf w_i|^2+\sigma_0^2}\right).
\end{equation}

The overall problem is to maximize some network utility {function ${\mathcal U}(R_1,\ldots,R_K)$} by optimizing the beamforming vectors at the BS $\{\mathbf{w}_k\}_{k=1}^K$ and the RIS
reflection coefficients $\mathbf{v}$. Now, since the channel coefficients are not known, 
we need to use a pilot training phase to learn the channel. Assuming 
a TDD system with channel reciprocity, we let each user $k$ send an uplink pilot 
sequence $x_k(\ell),$ $\ell=1,\ldots,L$, with $|x_k(\ell)|^2\le P_u$, to the BS. 
Then, the received pilots at the BS can be denoted as:
\begin{equation}\label{eq:uplink}
	\mathbf y(\ell) 
	=\sum_{k=1}^K\left(\mathbf h_k^{\rm d}+\mathbf A_k\mathbf{\tilde{v}}(\ell)\right)x_k(\ell)+\mathbf n(\ell),
\end{equation}
where $\mathbf{\tilde{v}}(\ell)$ is the vector of RIS reflection coefficients at the uplink transmission slot $\ell$ and can be thought of as part of the pilot, and
$\mathbf n(\ell)\sim\mathcal{CN}(\mathbf 0,\sigma_1^2\mathbf I)$ is the additive Gaussian noise. Denoting $\mathbf{Y} = [\mathbf y(1),\mathbf y(2),\cdots,\mathbf y(L)]\in\mathbb{C}^{M\times L}$ and $\mathbf W=[\mathbf w_1,\cdots,\mathbf w_k]$,
our goal is to design the downlink beamformers $\mathbf{W}$ and the reflection coefficients $\mathbf{v}$, based on the received uplink pilots $\mathbf{Y}$, which contains information about the channel.
This overall process can be thought of as solving the following optimization problem over the mappings
from $\mathbf{Y}$ to $(\mathbf{W}, \mathbf{v})$:
\begin{subequations}
\label{prob:formulation_ris}
\begin{align}
\underset{\begin{subarray}{c}
		(\mathbf W, \mathbf v)= \mathcal{G}(\mathbf{Y})
        \end{subarray}}{\maxi}\quad &\mathbb{E}\left[{\mathcal U}( R_1(\mathbf W,\mathbf v),\ldots, R_K(\mathbf W,\mathbf v) )\right] \\
\subj\quad& \displaystyle{\sum}_k\mathbb\|\mathbf w_k\|_2^2\le P_d,\\
        &|v_i| = 1,~~i=1,2,\ldots,N,
\end{align}
\end{subequations}
where the function $\mathcal{G}(\cdot):\mathbb{C}^{M\times L}\rightarrow \mathbb{C}^{M\times K}\times\mathbb{C}^{N}$ is the mathematical representation of the mapping to be optimized over, and the expectation is taken over the random channel realizations and the uplink noise.

Directly solving problem \eqref{prob:formulation_ris} is challenging, because it
involves optimizing over the high-dimensional mapping $\mathcal{G}(\cdot)$. 
The conventional approach is to first estimate the
channels from the received pilots $\mathbf Y$, then to solve the subsequent
network utility maximization problem based on the estimated channel. 
Instead, we propose a machine learning approach to directly learn such a mapping
using a GNN.


\begin{figure}[t]
    \centering
    \includegraphics[width=3.3in]{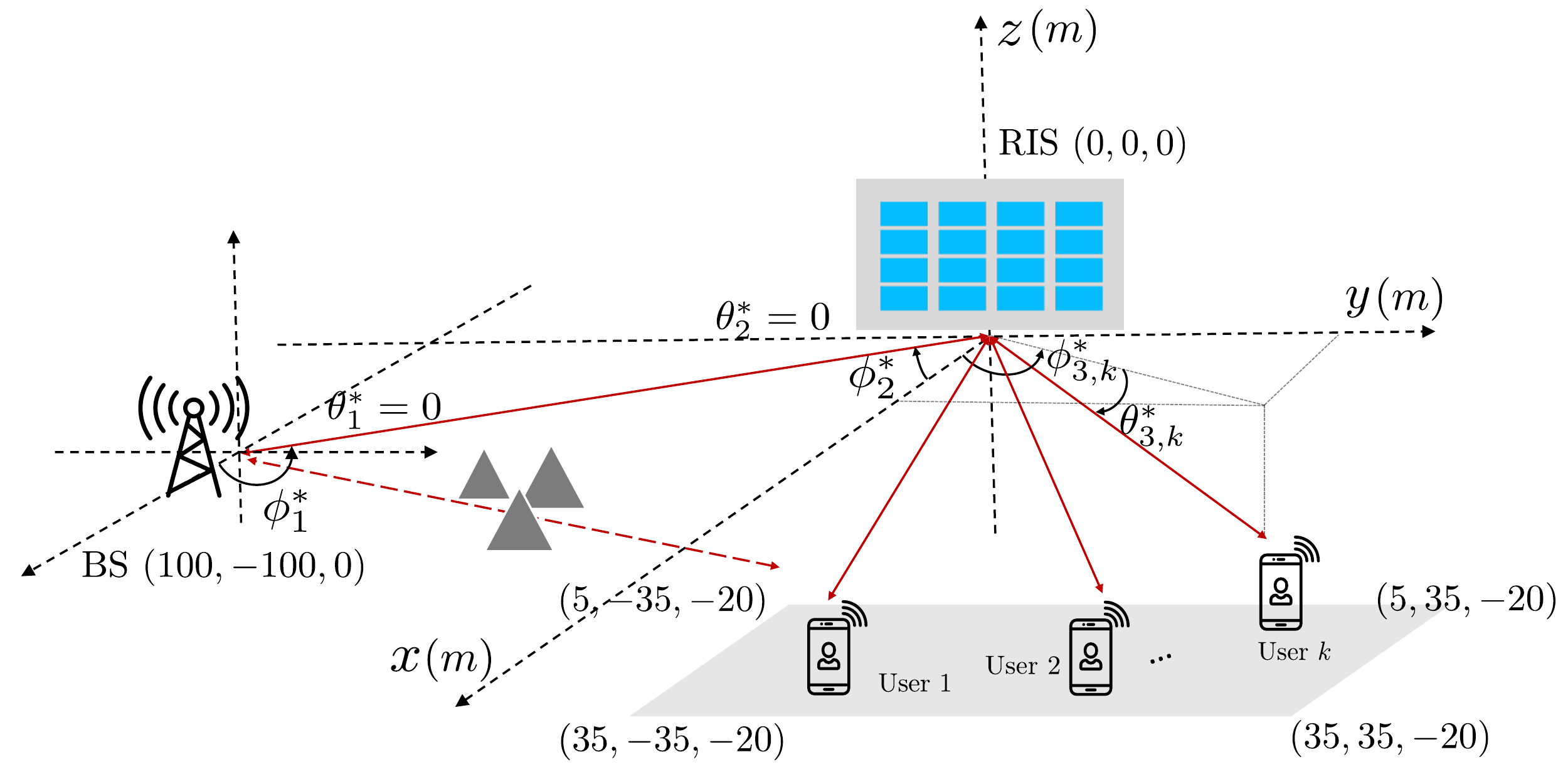}
	\caption{Geographic layout of an RIS-assisted downlink system. \cite{9427148}} 
    \label{fig_ris:simulation_layout}\vspace{0.3cm}
\end{figure}

\begin{figure}[t]
    \includegraphics[width=9.5cm]{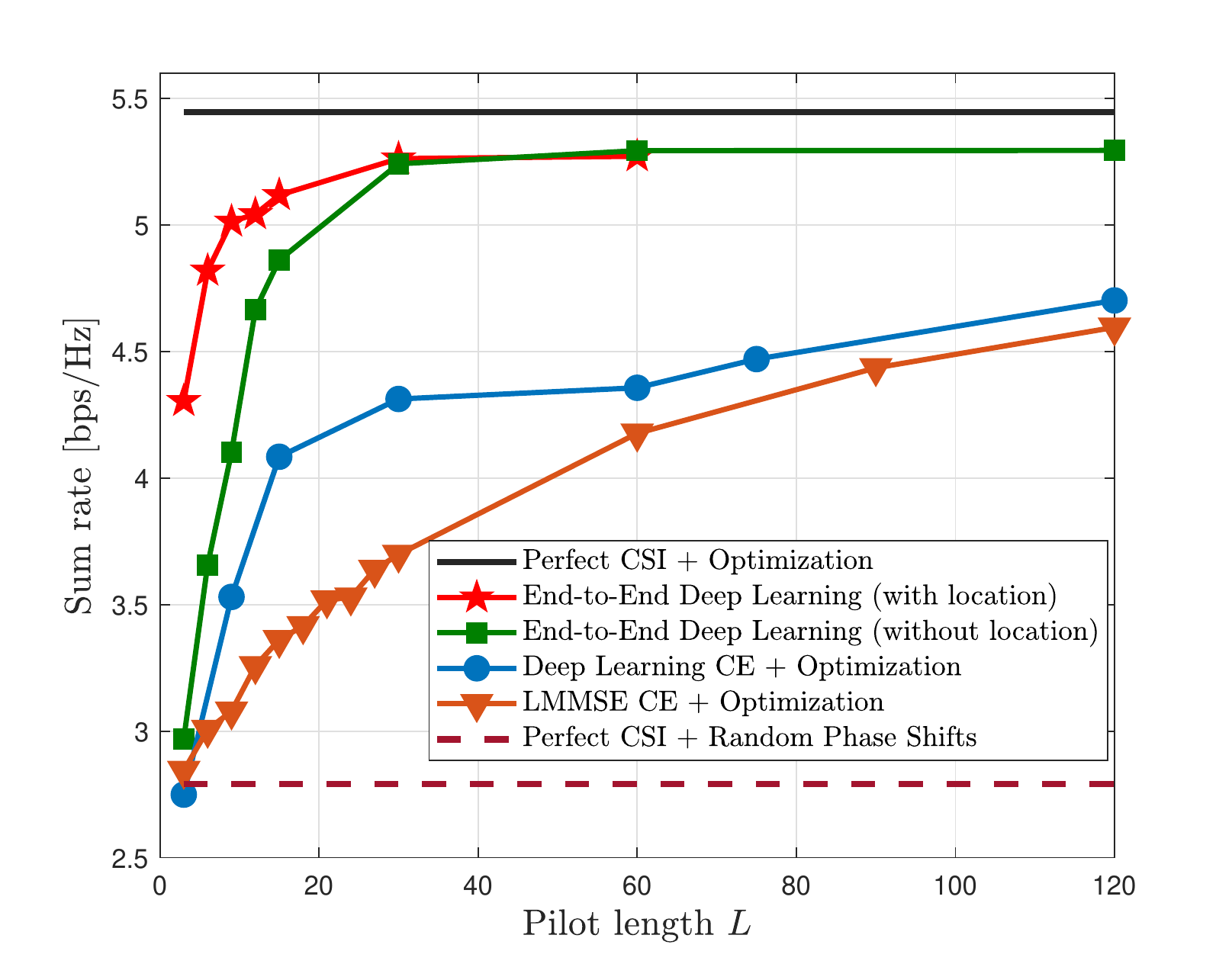}
	\caption{Sum rate versus pilot length for an RIS-assisted multiuser downlink system with an 8-antenna BS, a 100-element RIS, and 3 single-antenna users, comparing end-to-end deep learning approach to the conventional approach of channel estimation (CE) followed by RIS coefficients and BS beamforming optimization. \cite{9427148} } 
	\label{fig_ris:rate_vs_pilot}
\end{figure}

\begin{figure}[t]
    \centering
    \subfigure[Array response of the BS.]{ \includegraphics[width=7.7cm]{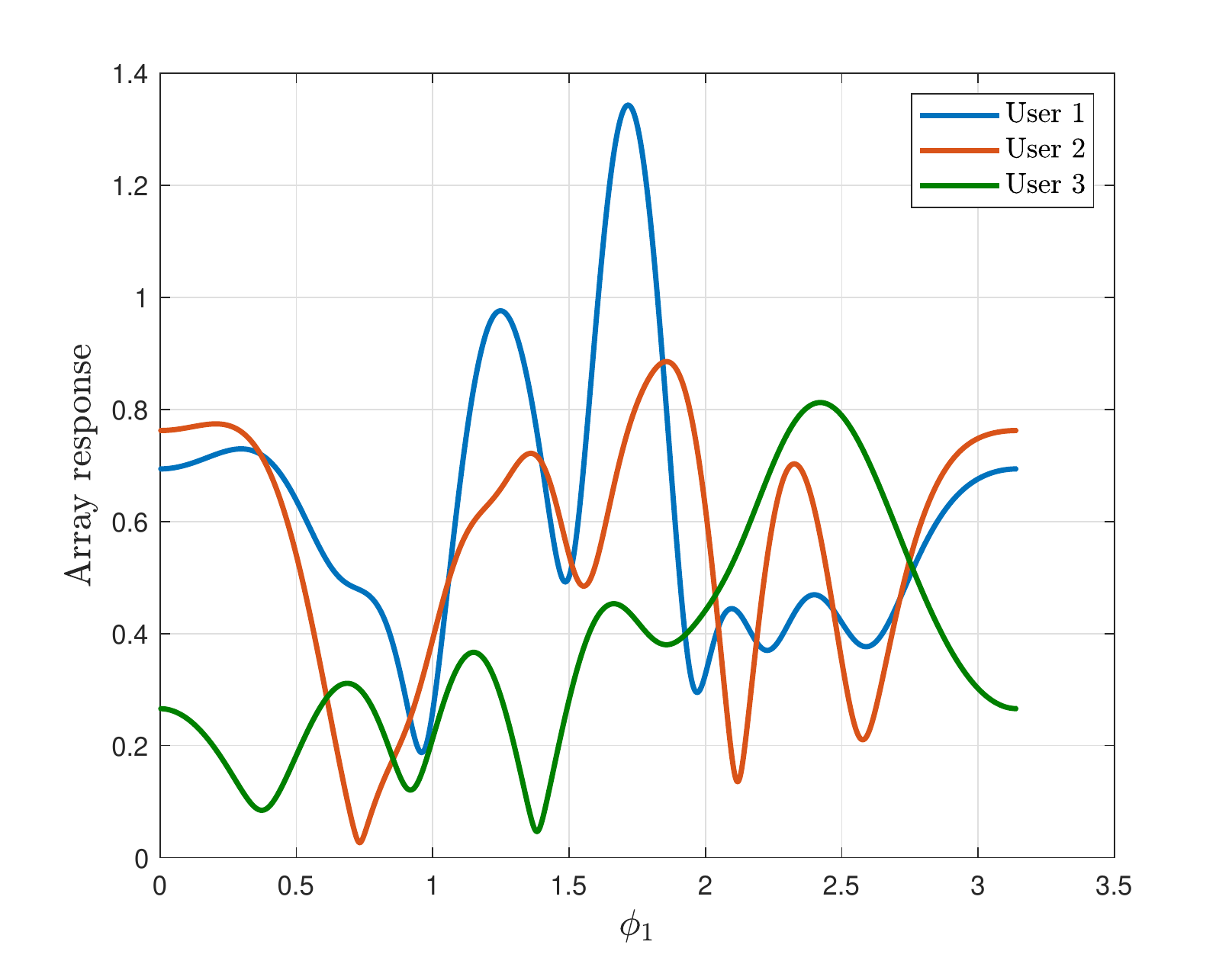}\label{fig_ris:bs_array_respons_mu}}
    \subfigure[Array response of the RIS.]{\includegraphics[width=7.8cm]{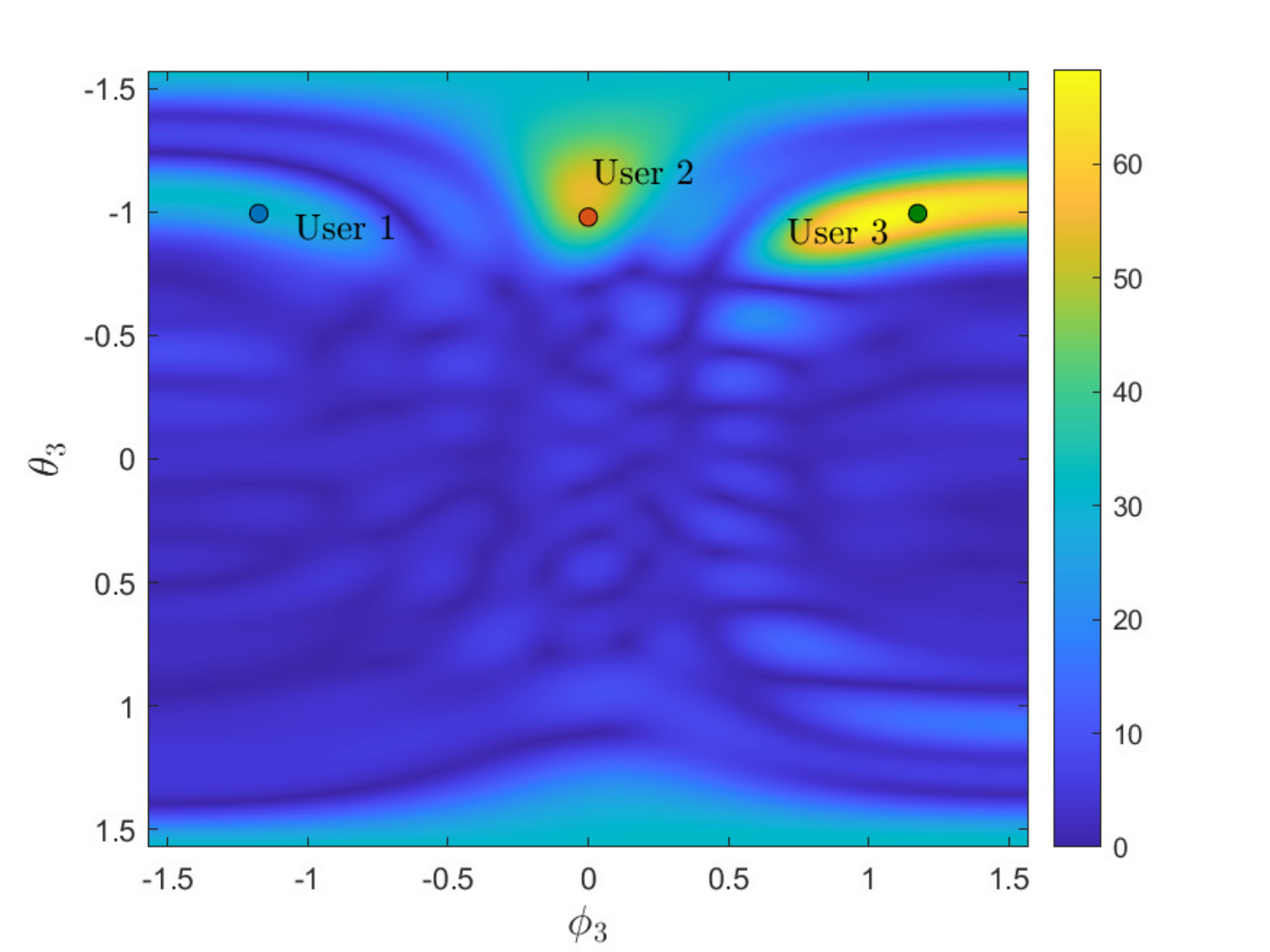}\label{fig_ris:irs_array_respons_mu}}
    \caption{The array response of the BS and the RIS obtained from the GNN for $N=100$ and $M=8$ for a 3-user system for maximizing the minimum rate.
	The true $(\phi_3^\ast,\theta_3^\ast)$ are $(-1.176, -0.994)$, $(0,-0.980)$, and $(1.176,-0.994)$, respectively, for the 3 users. \cite{9427148}}\label{fig_ris:array_response}
\end{figure}

\subsection{Learning to Beamform and to Reflect}

The overall learning framework is shown in Fig.~\ref{fig_ris:dnn_overall_arch},
where the received pilots after matched filtering {$\{\tilde{\bY}_k\}_{k=1}^K$}
is the input to a neural network that learns the optimized reflection coefficients
$\mathbf v$ and the beamforming matrix $\mathbf W$ without the intermediary
channel estimation step. 

The remaining key question is how to choose the neural network architecture. 
In theory, a fully connected neural network can already learn the mapping from 
the received pilots to the optimization variables.
However, a more efficient architecture is a one that captures the structure of 
the network utility maximization problem \eqref{prob:formulation_ris}. 
Specifically, observe that in \eqref{prob:formulation_ris}, if the indices of
users permute, the optimal RIS coefficients $\mathbf v$ should remain the same,
while the optimal beamforming vectors $\{\mathbf w_k\}_{k=1}^K$ should permute
in the same manner.  These properties are known as \emph{permutation invariance} and
\emph{permutation equivariance}. 

It is possible to design a neural network to automatically
enforce these properties.  This can be done using a GNN based on a graph
representation of the RIS and the users. The details of the GNN structure are
described in \cite{9427148}. The idea is to associate a representation vector $\mathbf{z}_k^d$ 
with each user and also with the RIS. The representation vectors are updated 
layer-by-layer, but the connections between the layers are based on aggregation and combination operations that are invariant with respect to input permutation, e.g., the ${\rm mean}()$ or $\max()$ functions.
After multiple layer iterations, the node representation vectors 
are mapped to the beamforming matrix $\mathbf W$ and 
the RIS coefficients $\mathbf{v}$. 
To make the architecture generalizable with respect to the number of users, the neural network weights across the users are tied together. 
The overall neural network can be trained to maximize the network utility function.

\subsection{Numerical Results}
To illustrate the performance of the machine learning approach for optimizing the 
beamformers and the reflective coefficients, we report the simulation results\footnote{The code for this simulation is available at \url{https://github.com/taojiang-github/GNN-IRS-Beamforming-Reflection}}
in \cite{9427148} on a scenario with $M=8$ antennas at the BS, $N=100$ elements at the RIS, and $3$ users. The direct-link channel $\mathbf{h}_k^{\rm d}$ is assumed to be Rayleigh fading, and the BS-RIS and RIS-users channels are assumed to be Rician fading with Rician factor set as $10$. The geographic locations of the BS, RIS, and users are shown in Fig.~\ref{fig_ris:simulation_layout}. 
The uplink pilot transmit power and the downlink data transmit power are respectively set to be $15$dBm and $20$dBm. The uplink and downlink noise power {are} $-100$dBm and $-85$dBm, respectively. 

Fig.~\ref{fig_ris:rate_vs_pilot} plots the average sum rate versus pilot length for different approaches. As can be seen from Fig.~\ref{fig_ris:rate_vs_pilot}, the performance of the {linear minimum mean-squared-error (LMMSE)} channel estimation based method is able to 
approach the perfect channel state information (CSI) baseline as the pilot length increases. However, the end-to-end deep learning method approaches the perfect CSI baseline much faster, showing that the GNN can utilize the pilots in a more efficient way. 

We also provide the simulation results on the model-then-optimize approach in which a GNN is used for explicit channel estimation, and the beamforming matrix and RIS coefficients are optimized based on the estimated channel. While this method shows better performance as compared to the LMMSE based approach, its performance is still much worse than the GNN approach that directly learns the solution from the pilots. This shows the benefit of bypassing explicit channel estimation. 
Moreover, additional information such as the locations of the users can be easily incorporated in the end-to-end deep learning framework, which can further improve the performance as shown in Fig.~\ref{fig_ris:rate_vs_pilot}.

The GNN produces interpretable solutions.
Fig.~\ref{fig_ris:array_response} shows the array responses 
learned by the GNN for a maximizing minimum rate problem for three users 
at different locations. 
We observe from Fig.~\ref{fig_ris:irs_array_respons_mu} that the learned RIS
coefficients indeed focus the beams to the corresponding user locations, but
the three users get different focusing strengths. Interestingly, because the BS
beamformers and the RIS reflective coefficients are designed jointly, the user
corresponding to the weakest RIS focusing is compensated by a
stronger BS beamforming gain as seen in Fig.~\ref{fig_ris:bs_array_respons_mu}. 
Thus, the combined channel strengths are equalized across the three users. 
Overall, these results show that the GNN indeed is able to learn
interpretable solutions, based on much fewer pilots than the conventional strategies.

\section{Distributed Source Coding for Channel Estimation and Feedback in FDD Massive MIMO}
\label{sec:CSI_feedback}

\begin{figure}[t]
\centering
    \includegraphics[width=3.3in]{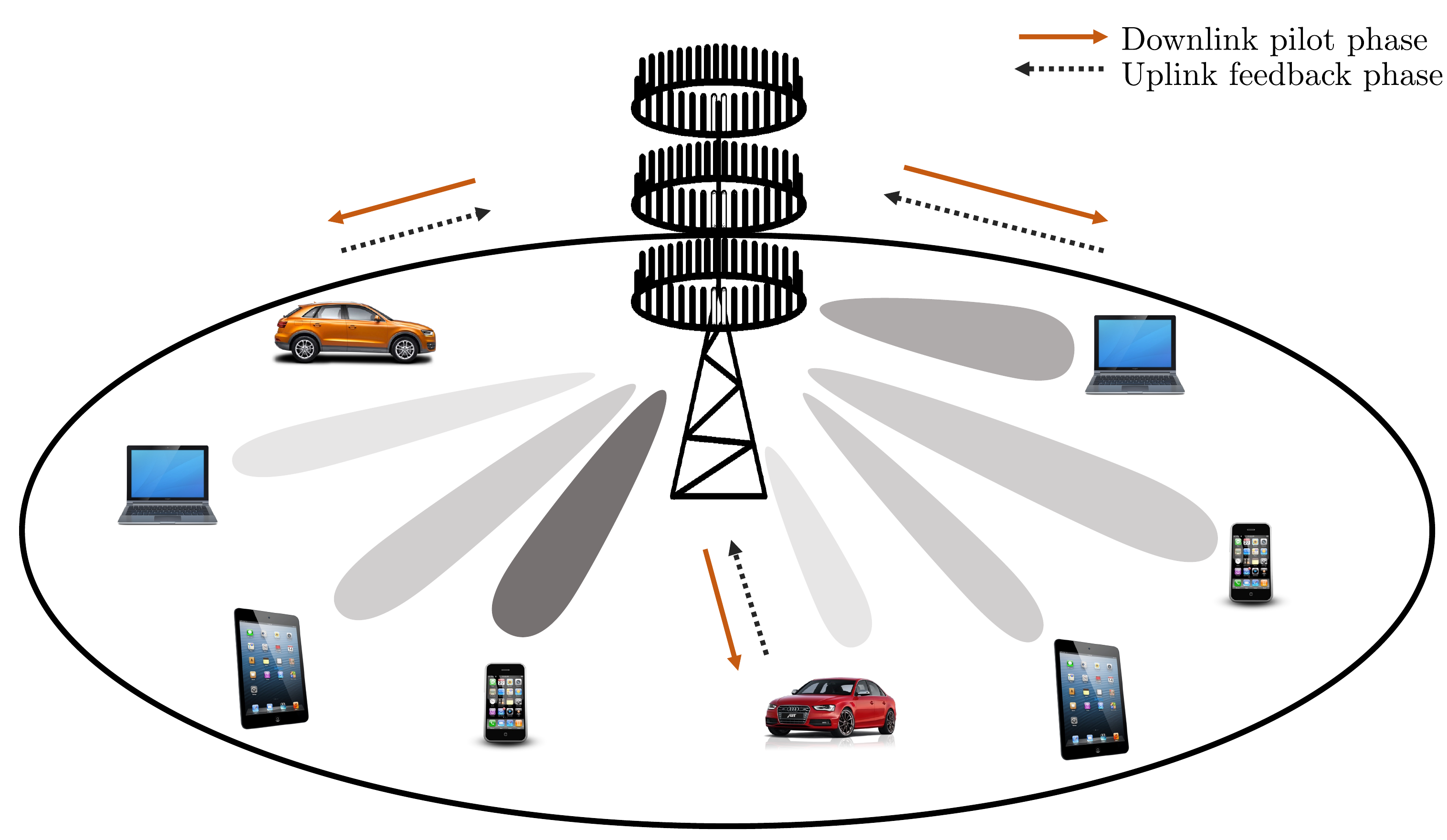}
    \caption{An FDD massive MIMO system, in which the BS transmits the pilots, the users estimate their channels then feedback a quantized version of the channels to BS, and the BS designs the precoders based on the feedback from all the users. \cite{9347820}}\label{fig_fdd:system}
\end{figure}

The channel estimation problem is more challenging in the frequency-division
duplex (FDD) system, which cannot rely on channel reciprocity. In this case, 
as shown in Fig.~\ref{fig_fdd:system}, the pilots are sent by the BS in the
downlink and are observed by the users.  The users need to estimate their
channels, then send quantized versions of the channels through rate-limited feedback links to the BS, so that the BS can design a
precoding strategy to serve all the users.  The conventional approach to this
problem relies on model-based channel estimation followed by independent
codebook-based quantization and feedback \cite{Love2008,Gao2019,Rao2014}. This is far from optimal. We show
here that machine learning techniques can be used to train a set of optimized
distributed source encoders together with a centralized decoder in an
end-to-end fashion in order to maximize a system-level objective.  Such an
approach can significantly reduce the length of pilots needed to achieve the
maximum throughput in an FDD massive MIMO system.

The channel estimation and feedback design for a multiuser FDD massive MIMO 
system can be thought of as a distributed source coding problem. 
Distributed source coding is a long-standing information 
theoretical problem in which distributed encoders compress their observations 
for centralized reconstruction at the decoder. Here, the users are the
distributed source encoders who observe then quantize a noisy version of the
sources. The BS is the centralized source decoder, which aims to compute a 
function of the sources.

The optimal design of distributed source encoders and decoder is highly nontrivial. Information theoretic optimal coding strategies involve concepts such as binning, which can be thought of as a multiuser codebook. While it is unlikely for a neural network to learn structured binning, it can help design good codebook-based quantization and feedback strategies that reap the benefit of distributed source coding. This is an example in which a data-driven approach can play an important role in designing short block-length quantization codes under rate constraints.

\begin{figure*}[t]
   \centering
   \subfigure[]
{ \includegraphics[width=7.8cm]{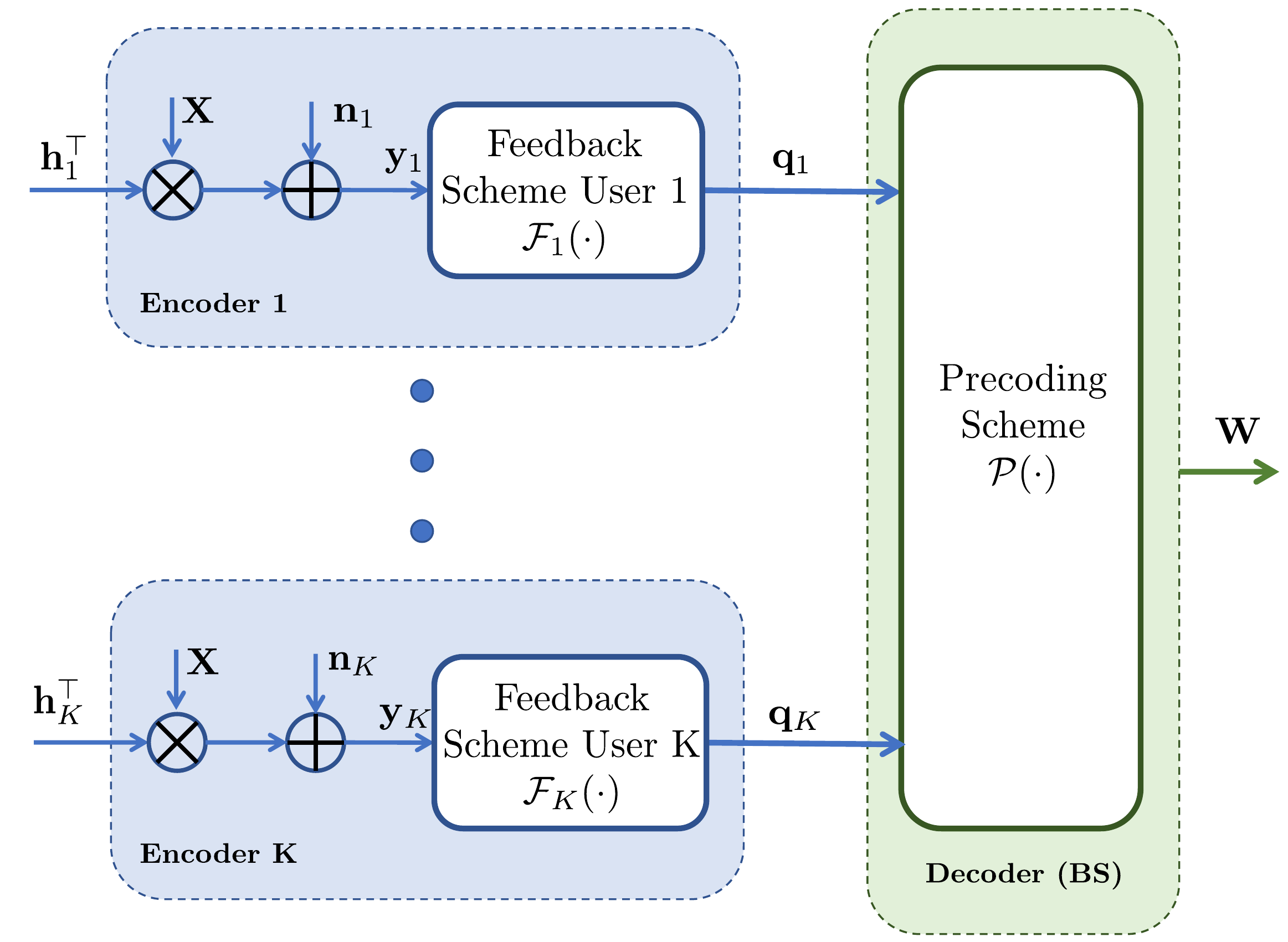}\label{fig_fdd:fdd_a}}\hspace{0.2cm}
   \subfigure[]
{\includegraphics[width=9.2cm]{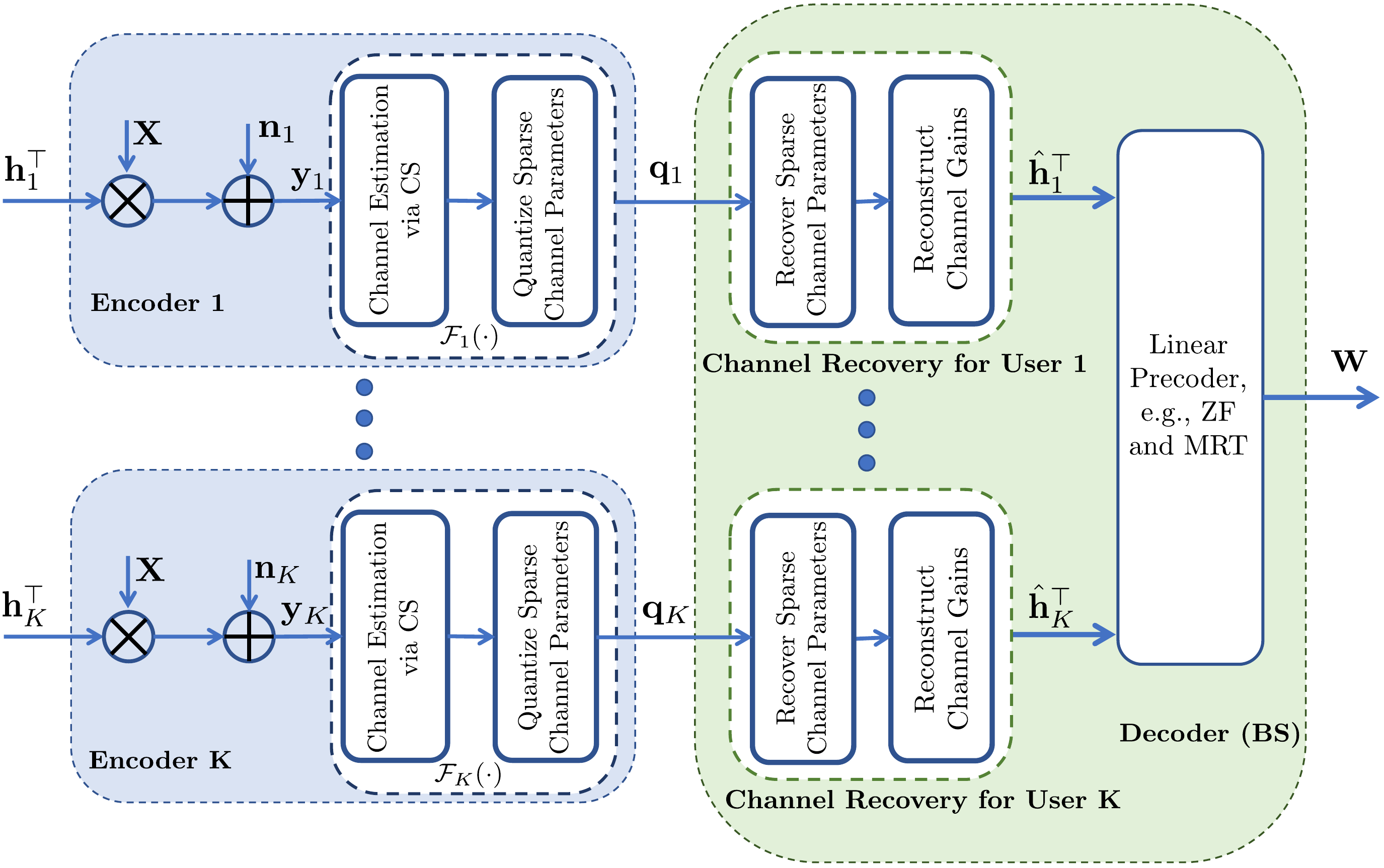}\label{fig_fdd:fdd_b}}
   \caption{Comparison between end-to-end design and conventional scheme in FDD downlink precoding problem. 
(a) The FDD downlink precoding design problem can be viewed as a distributed source coding problem in which the downlink pilots and the feedback schemes adopted at the users can be thought of as the source encoders and the precoding scheme adopted at the BS can be thought of as the decoder;
(b) The conventional channel feedback scheme can be regarded as a separate source coding strategy of independent quantization of each user's channel.
In the machine learning approach, the feedback scheme at the user side and the precoding scheme at the BS side are replaced by DNNs that can be trained in an end-to-end fashion.  \cite{9347820} }\label{fig_fdd:fig_fdd}
\end{figure*}

\subsection{System Model and Problem Formulation}
Consider an FDD multiuser MIMO system in which a BS equipped with $M$ antennas serves $K$ single-antenna users. Analogues to the previous section, we consider the downlink scenario in which the BS aims to communicate the data symbol $s_k \in \mathbb{C}$ with $\mathbb{E}[|s_k|^2]=1$ to user $k$ using a precoding vector $\mathbf w_k\in\mathbb{C}^M$, which satisfies a total power constraint $\sum_{k=1}^K\mathbb\|\mathbf w_k\|_2^2\le P_d$. Assuming a narrowband block-fading channel model, the received signal at the $k$-th user in the data transmission phase can be written as:
\begin{equation}\label{eq_rx_sig}
r_k =  \bh_k^\top \bw_k s_k + \sum_{i\not=k} \bh_k^\top \bw_ i s_i + z_k,
\end{equation}
where $\bh_k \in \mathbb{C}^M$ is the channel between the BS and user $k$ and $z_k \sim \mathcal{CN}(0,\sigma_0^2)$ is the additive white Gaussian noise. 
The achievable rate of user $k$ is given by:
\begin{equation}\label{eq:ratek}
    R_k = \log_2\left(1 +  \frac{\lvert \bh_k^\top \bw_k \rvert^2}{ \sum_{i\not=k} \lvert \bh_k^\top \bw_i \rvert^2+\sigma_0^2 } \right).
\end{equation}



The aim is to maximize a network utility function $\mathcal{U}(R_1,\ldots,R_K)$, which is a function of the precoding vectors $\{\bw_k\}_{k=1}^{K}$. To design the optimal precoding vectors, the BS must first acquire the instantaneous CSI. 
We consider a pilot phase for the FDD system in which the BS sends pilots ${\bX}\in \mathbb{C}^{M\times L}$ of length $L$, and the $k$-th user receives ${\by}_k\in\mathbb{C}^{1\times L}$ as
\begin{equation}\label{eq_rx_pilot}
    {\by}_k = \bh_k^\top {\bX} + {\bn}_k,
\end{equation}
where the pilots in the $\ell$-th transmission satisfy the power constraint, i.e., $\|\bx_\ell\|_2^2\leq P_d$ with $\bx_\ell$ being the $\ell$-th column of $\bX$, and ${\bn}_k \sim \mathcal{CN}(\mathbf{0},\sigma_0^2\mathbf{I})$ is the additive white Gaussian noise at user $k$. Subsequently, the $k$-th  user abstracts the useful information in the received pilots $\by_k$ for the purpose of multiuser downlink precoding, and feeds back that information to 
the BS under a feedback constraint of $B$ bits, i.e., 
\begin{equation}\label{eq_feedback}
\bq_k = \mathcal{F}_k\left( {\by}_k\right), 
\end{equation}
where the function $\mathcal{F}_k: \mathbb{C}^{1\times L} \rightarrow \{\pm 1\}^B $ is the $k$-th user feedback scheme. Finally, the BS designs the multiuser precoding matrix  $\mathbf W=[\mathbf w_1,\cdots,\mathbf w_k]$ based on the feedback bits received from all $K$ users (i.e., $\bq = [\bq_1^\top,\bq_2^\top,\ldots,\bq_K^\top]^\top$), i.e., 
   \begin{equation}\label{eq_decoding}
\bW = \mathcal{P} \left( \bq \right), 
\end{equation}
where the function $\mathcal{P}: \{\pm 1\}^{KB} \rightarrow   \mathbb{C}^{M\times K} $ denotes the multiuser downlink precoding scheme.

The overall problem formulation is therefore 
\begin{subequations}
\begin{align}
\displaystyle{\Maximize_{{\bX},\hspace{2pt}\{\mathcal{F}_k(\cdot)\}_{\forall k},\hspace{2pt}\mathcal{P}(\cdot)}} ~~ & \mathcal{U}(R_1,\ldots,R_K) \\ 
\text{subject to}  \quad~~ & \bW = \mathcal{P}\left(\left[\bq_1^\top,\ldots, \bq_K^\top \right]^\top \right),\\
~&\bq_k =  \mathcal{F}_k(\bh_k^\top {\bX} + {\bn}_k), ~~\forall k, \\
~& \displaystyle{\sum}_k \mathbb\|\mathbf w_k\|_2^2\le P_d,\\
~&\|\bx_\ell\|^2_2\leq P_d, ~~\forall \ell,
\end{align}
\label{main_problem}
\end{subequations}
in which the training pilots ${\bX}$, all $K$ users' feedback schemes $\{\mathcal{F}_k(\cdot)\}_{k=1}^{K}$, and the multiuser precoding scheme $\mathcal{P}(\cdot)$ can be designed to optimize the overall utility function of the system.

This problem can be viewed as a distributed source coding problem with the network
utility as the ``distortion'' metric, 
because channel estimation and quantization are performed across $K$
distributed users, and the feedback bits from all $K$ users are centrally
processed at the BS for the purpose of designing the multiuser precoder,
as illustrated in Fig.~\ref{fig_fdd:fdd_a}. 
Obtaining the optimal distributed source coding strategy by directly solving the optimization problem
\eqref{main_problem} is challenging. As shown in
Fig.~\ref{fig_fdd:fdd_b}, the conventional design of FDD massive
MIMO system is based on independent quantization and
feedback of the channel vector (or channel parameters) at each user. However,
such independent quantization and feedback approach is
quite suboptimal, especially in the short pilot regime. 
In this section, we show that a deep learning approach can be used to design
a more efficient distributed source coding codebook 
for the FDD massive MIMO systems.

\subsection{Learning Distributed Channel Estimation and Feedback}

The idea is to use DNNs to model the feedback scheme $\{\mathcal{F}_k(\cdot)\}_{k=1}^{K}$ 
and the multiuser precoding scheme $\mathcal{P}(\cdot)$ in Fig.~\ref{fig_fdd:fdd_a}. 
The rest of this subsection briefly explains how we solve the overall optimization 
problem \eqref{main_problem} by employing such a deep learning framework.

As the first step of the downlink training phase, the BS sends $L$ training pilots and the $k$-th user observes the pilots through its channel as ${\by}_k = \bh_k^\top {\bX} + {\bn}_k$. Since the received signal $\by_k$ is a linear function of the channel $\bh_k$, we can simply model it as the output of a single-layer neural network with linear activation function in which the input is the channel $\bh_k$. In this single-layer neural network, the weight matrix is the pilot ${\bX}$ and the bias vector is the noise vector $\mathbf{n}_k$. 
To enforce the total power constraint on each pilot transmission, we adopt a weight constraint under which each column of ${\bX}$ satisfies $\|\bx_\ell\|^2_2 \le P_d$. It is worth mentioning that such a weight constraint is often used in the machine learning literature for regularization in order to reduce overfitting, e.g., \cite{hinton2012improving}. Here, we use the weight constraint to capture the physical constraint on the downlink power level of transmit antennas of a cellular BS.

At the user side, upon receiving ${\by}_k$, the user seeks to summarize the useful information in $\by_k$ and to feed back that information to the BS in a form of $B$ information bits. We can simply model this process by a DNN which maps $\by_k$ to feedback bits $\bq_k$. To make sure that the final output of the DNN is in the form of binary bits, we use the sign activation function at the last layer of the user-side DNNs.

Finally, assuming an error-free feedback channel between each user and the BS,
the BS designs precoding vectors as a function of the received feedback bits from all $K$ users. We propose to use another DNN to map the received feedback bits $\bq$ to the design of the multiuser precoding matrix $\bW$. To ensure that the precoding matrix designed by the DNN satisfies the total power constraint, we employ a normalization layer at the last layer of the BS-side DNN.

The overall distributed source coding strategy is designed by training the
end-to-end deep learning framework to maximize the network utility using 
stochastic gradient descent. 
But care must be taken, due to the fact that the derivative of the sign activation
function is always zero, so the conventional back-propagation method cannot be
directly used to train the overall network. It is possible to circumvent 
this difficulty by adopting the straight-through approximation
in which the sign activation function is approximated by another smooth
differentiable function for the back-propagation step \cite{chung2016}. 
By gradually tightening the approximation, we eventually arrive at a beamforming
codebook that maps the noisy version of the channels from all the users to an optimized
set of downlink beamformers.

\begin{figure}[t]
        \centering
        \includegraphics[width=9.5cm]{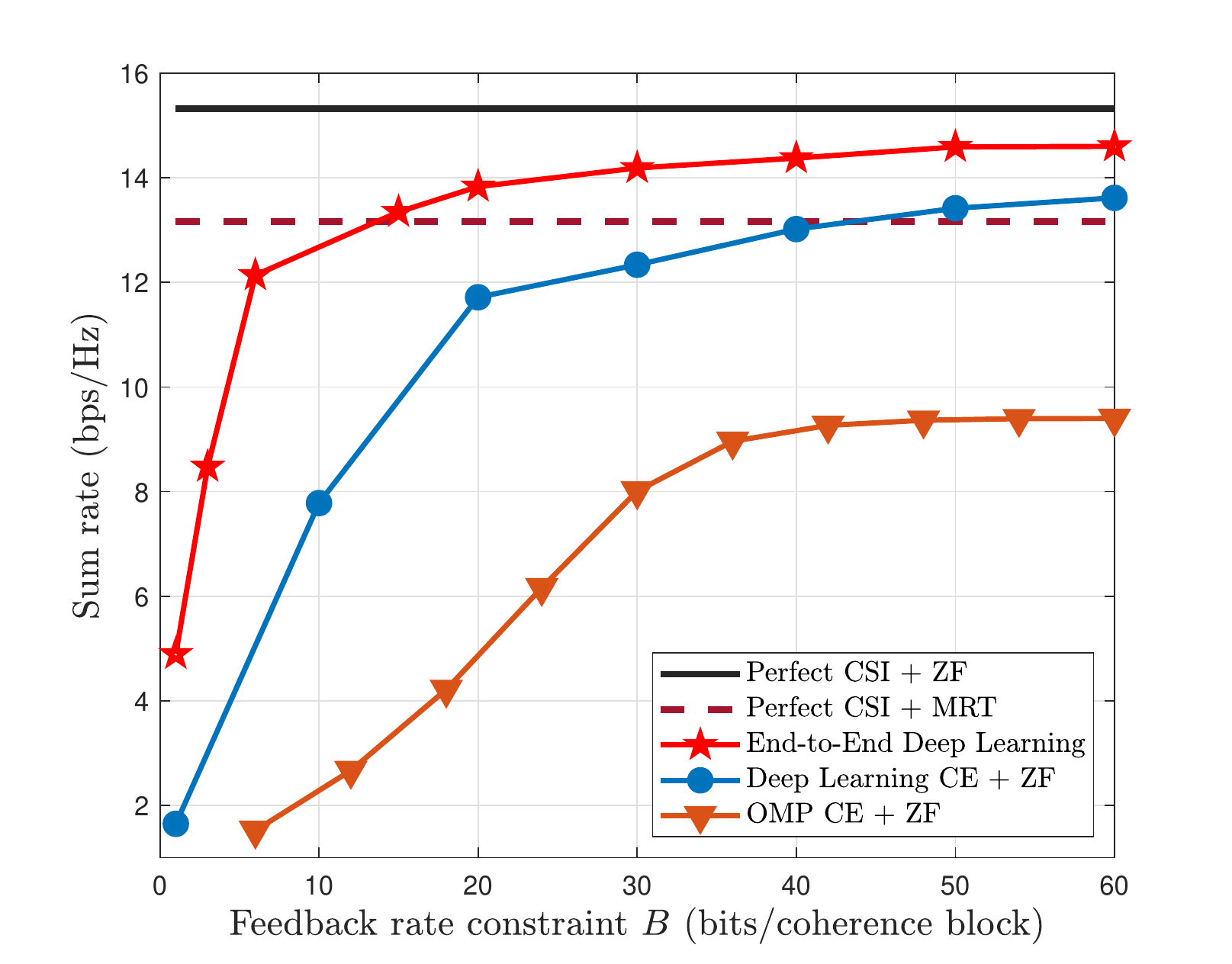}
        \caption{Sum rate versus feedback rate constraint in a 2-user FDD massive MIMO system with $64$ antennas, sparse channels with 2 dominant paths, pilot length of $8$ and SNR of 10dB.  \cite{9347820} }
       \label{fig:L8}
\end{figure}

\subsection{Numerical Results}
We now present the performance evaluation of the end-to-end deep learning framework in a scenario where a BS with $M=64$ antennas serves $K = 2$ users in a mmWave propagation environment with $2$ dominant paths as reported in \cite{9347820}\footnote{The code for this simulation is available at \url{https://github.com/foadsohrabi/DL-DSC-FDD-Massive-MIMO}}. The fading coefficient of each path is modelled by a Gaussian random variable and the corresponding angle of departure is modelled by a uniform random variable in the range of $[-30^\circ,30^\circ]$. The signal-to-noise ratio (SNR) $P_d/\sigma_0^2$ is set to 
$10$dB and the pilot length $L=8$.   

Fig.~\ref{fig:L8} plots the average sum rate versus per-user feedback rate constraint $B$. It can be seen that the end-to-end deep learning framework with relatively low rate feedback links (i.e., about $15$ bits per user) can already outperform the maximum-ratio transmission (MRT) precoding baseline with full CSI. The MRT precoding design does not take the inter-user interference into account. This shows that the trained DNN has actually learned a precoding mechanism capable of alleviating inter-user interference in a multiuser FDD massive MIMO system.

Furthermore, we compare the performance of the end-to-end deep learning framework with that of the conventional design methodology based on channel estimation followed by linear precoding schemes such as zero forcing (ZF). For the channel estimation part of the conventional approach, two different methods are used: (i) a compressed sensing algorithm called orthogonal matching pursuit (OMP) and (ii) deep learning-based channel estimation method. 

Fig.~\ref{fig:L8} shows that the end-to-end deep learning framework can achieve a significantly better performance as compared to the conventional channel estimation based design methodology (either when the channel estimation is implemented by OMP or by deep learning). This confirms the intuition that in practical massive MIMO systems in which the pilot length is much smaller than the number of antennas, the conventional approach of first estimating then quantizing the sparse channel parameters is quite suboptimal. 
The end-to-end deep learning framework can achieve much better performance, because it is able to better explore the channel sparsity. It implicitly estimates the channel and designs the quantization codebooks jointly across the multiple users 
in order to maximize an overall true system objective, i.e., the sum rate in this case.

\section{Active Sensing for mmWave Channel \\ Initial Alignment}
\label{sec:active_sensing}

Machine learning also has an important role to play in solving high-dimensional nonconvex optimization problems in sensing applications. To illustrate this point, we consider the mmWave initial alignment problem for a BS equipped with a
hybrid massive MIMO architecture, consisted of an analog beamformer and a
low-dimensional digital beamformer. The user transmits a sequence of pilot signals; 
the BS makes a corresponding sequence of observations, via the analog beamformers, which it can
design, but the observations reside only in the low-dimensional digital domain.
The question is in which analog directions should the BS choose to observe in a sequential manner 
in order to obtain the most accurate channel information for a communication or sensing task of interest?

Because the sensing direction in each stage can be designed as
a function of the previous observations, this is an active sensing problem for which 
the analytic solution is highly nontrivial and the conventional codebook-based approach is
highly suboptimal \cite{alkhateeb2014channel,Tara2019Active}. Specifically, \cite{alkhateeb2014channel} proposes a bisection search algorithm to gradually narrow down the AoA range. However, the performance of the bisection algorithm is very sensitive to noise power, 
so it is suitable for the high SNR scenario only.
To address this issue, \cite{Tara2019Active} proposes to select the next sensing vector from a predefined codebook based on the posterior distribution of the angle-of-arrival (AoA). Further, \cite{9448070} eliminates the codebook constraint by directly mapping the posterior distribution to the next sensing vector using a DNN. However, as the computation of posterior distribution is applicable only to the single-path channel model, the generalization of these ideas to the multipath channel is challenging.

Instead, we show that an excellent solution can be obtained by
training a DNN to learn the sensing direction in an end-to-end manner without needing 
to compute the posterior.
Further, we explore the active nature of the problem and show that by using a
long short-term memory (LSTM) based architecture \cite{lstm}, the state representation
in each observation stage can be learned and be used to design the sensing
direction in the next stage. The results show that machine learning can offer
a significant advantage over the current state-of-the-art. 

\subsection{System Model and Problem Formulation}
Consider a TDD mmWave system in which a BS equipped with $M$ antennas and a single RF chain serves a single-antenna user. The user transmits a sequence of pilots to the BS, and the BS seeks to estimate the channel or to design a subsequent downlink beamformer to maximize the beamforming gain, based on the received pilots. Due to the limited RF chain, the BS can only sense the channel through an analog beamformer (or combiner), but it can design the analog beamformers sequentially to sense different directions over time. Specifically, in time frame $t\in\{1,\ldots, T\}$, let $\mathbf w_t\in\mathbb{C}^M$ denote the sensing (i.e., combining) vector with $\|\mathbf w_t\|_2^2=1$ and let $x_t=\sqrt{P_u}$ be the pilot symbol, then the received pilot at the BS is given by:
\begin{align}
    y_t = \mathbf{w}_t^{\top} \mathbf{h}x_t+n_t,
\end{align}
where $n_t\sim\mathcal{CN}(0,\sigma_1^2)$ is the effective noise, and $\mathbf{h}\in\mathbb{C}^{M}$ is the channel from the user to the BS. In a mmWave environment, the channel $\mathbf{h}$ is often sparse, and can typically be modelled in the form of a multipath channel as follows:
\begin{equation}
    \mathbf{h} = \sum_{i=1}^{L_{\rm p}}\alpha_i \mathbf{a}_i({\phi}_i), 
\end{equation}
 where $L_{\rm p}$ is the number of paths, $\alpha_i\in\mathcal{CN}(0,1)$ is the fading coefficient of the $i$-th path, and $\phi_i\in[\phi_{\min},\phi_{\max}] $ is the AoA of the $i$-th path, and
$\mathbf{a}(\phi) = \left [ 1, e^{j {\pi } \sin{\phi} },..., e^{j(M-1){\pi } \sin{\phi}}  \right]^\top$ is the array response vector.

\begin{figure}[t]
        \centering
        {\includegraphics[width=\linewidth]{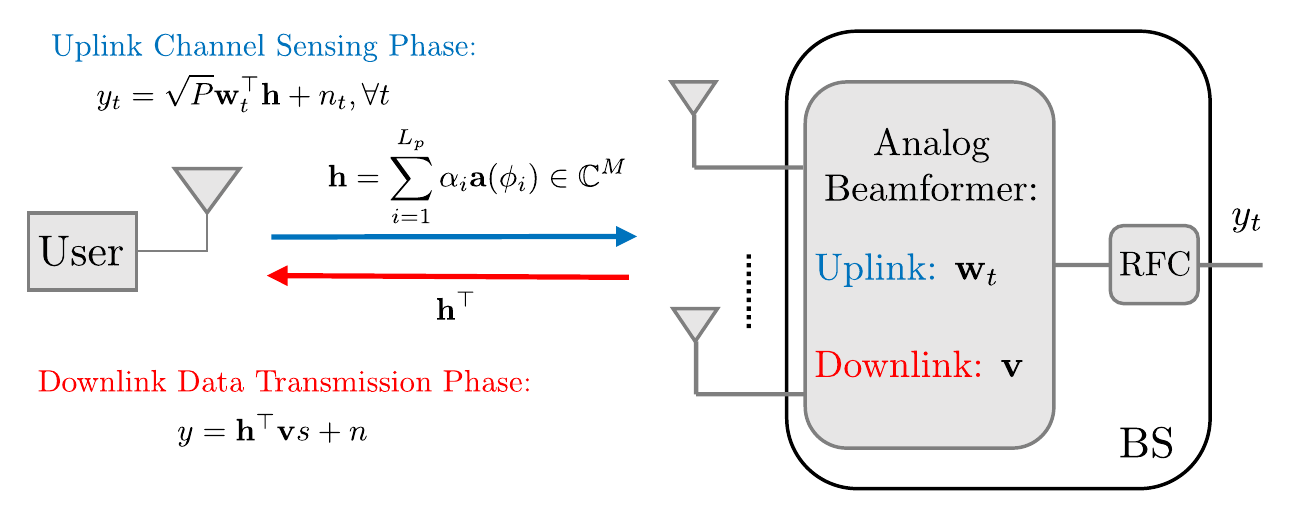}}
	\caption{Active sensing for mmWave initial alignment at a BS with a single RF chain. The goal is to design the analog sensing beamformers $\mathbf{w}_t$ adaptively as a function of the previous observations over multiple sensing stages $t=1,\cdots,T$ for the purpose of maximizing a utility function, e.g., the eventual downlink transmission beamforming gain $|\mathbf{h}^\top \mathbf v|^2$ after the sensing stage.  \cite{9724252}  }
	\label{Fig_sys_model_active}
\end{figure}

\begin{figure*}[t]
 \centering
\includegraphics[width=0.86\textwidth]{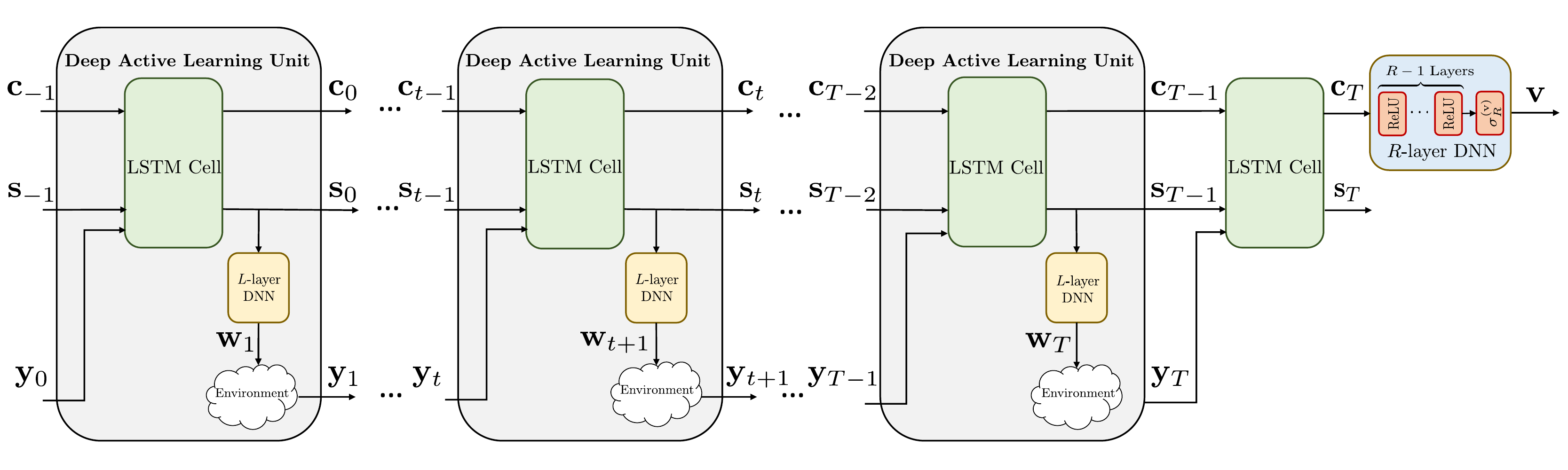}
\caption{An LSTM-based sequential learning architecture for solving an active sensing problem in mmWave initial alignment, in which the LSTM cells aim to summarize the system state based on the observations made so far, and a DNN is used to produce the analog combiner to be used in the next sensing stage.  \cite{9724252} }
\label{fig:active_learning_entire}
\end{figure*} 

Assuming a fixed total number of pilot stages $T$, the objective in the active sensing problem is to sequentially design the sensing beamformers $\{\mathbf{w}_t\}_{t=1}^T$ to maximize some utility function, i.e., $\mathcal{J}(\boldsymbol{\theta},\mathbf{v})$,  where $\boldsymbol{\theta}=\{\alpha_i,\phi_i\}_{i=1}^{L_{\rm p}}$ contains all the channel parameters and $\mathbf{v}\in\mathbb{C}^V$ is the
parameter to be designed or estimated after receiving all the pilots. 
For example, as illustrated in Fig.~\ref{Fig_sys_model_active}, $\mathbf{v}\in\mathbb{C}^M$ can be the subsequent downlink data transmission beamformer and the goal can be to maximize the beamforming gain, i.e., $\mathcal{J}(\boldsymbol{\theta},\mathbf{v}) := |\mathbf{h}^\top \mathbf v|^2$. In other applications, e.g., AoA-based localization, we might be interested in estimating the AoAs of the multipath channel, i.e., $\mathcal{J}(\boldsymbol{\theta},\mathbf{v}) := -\sum_{i=1}^{L_{\rm p}}(\hat{\phi}_i-\phi_i)^2$, where $\mathbf{v} := [\hat{\phi}_1,\cdots,\hat{\phi}_{L_{\rm p}}]^\top$. The key characteristic of such problems is that the sensing vector $\mathbf{w}_{t+1}$ can be designed based on the historical observations at stage $t$. Accordingly, the overall problem can be formulated as: 
\begin{subequations}
\label{eq:problem_formulation_unsup}
\begin{align}
\Maximize_{\left\{\mathcal{G}_t(\cdot,\cdot)\right\}_{t=0}^{T-1},\hspace{1pt} \mathcal{F}(\cdot,\cdot) }& \mathbb{E}\left[ \mathcal{J}(\boldsymbol{\theta} ,\bv) \right]\\
\text{subject to}\hspace{14pt} &\bw_{t+1} = \mathcal{G}_t\left(\by_{1:t},\bw_{1:t}\right),~ t=0,\ldots,T-1,\\
& \bv = \mathcal{F}\left(\by_{1:T},\bw_{1:T}\right),
\end{align}
\end{subequations}
where $\mathcal{G}_t:  \mathbb{R}^{t} \times \mathbb{R}^{tM} \rightarrow \mathbb{R}^M$ is the adaptive sensing strategy adopted by the BS in time frame $t$ and $\mathcal{F}: \mathbb{R}^{T} \times \mathbb{R}^{TM} \rightarrow \mathbb{R}^V$ is the function for designing the vector $\mathbf{v}$.

The active sensing problem \eqref{eq:problem_formulation_unsup} is challenging
to solve, because both the active sensing strategy
$\left\{\mathcal{G}_t(\cdot,\cdot)\right\}_{t=0}^{T-1}$ and the mapping
$\mathcal{F}(\cdot,\cdot)$ are functions in high-dimensional spaces. 
Moreover, the input dimension of the function
$\mathcal{G}_t(\cdot,\cdot)$ increases as the number of sensing stages
increases, making the sensing strategy particularly difficult to design when
$T$ is large. The conventional strategies are codebook based. For example,
a hierarchical beamforming codebook \cite{alkhateeb2014channel} can be designed 
based on the principle of bisection as mentioned before. A posterior matching 
based approach for sequentially selecting the appropriate analog combiners from 
the hierarchical codebook is proposed in \cite{Tara2019Active}. But these approaches 
are by no means optimal and are restricted to single-path channels. For the
multipath channel, nonadaptive sensing strategies which exploit the channel
sparsity are usually adopted \cite{alkhateeb2014channel}. 

In this section, we show that instead of using a model-based approach, a
codebook-free data-driven approach can be used to design the analog
combiners to sense a multipath channel.  Specifically, the sequential nature of
the problem suggests that a recurrent neural network (RNN) is an appropriate
network architecture. We show that a deep active sensing framework based on
the LSTM network, which is a variation of RNN, can be used to efficiently solve
the active sensing problem~\eqref{eq:problem_formulation_unsup}.


\subsection{Learning Active Sensing Strategy}

The proposed active sensing framework is as shown in Fig.~\ref{fig:active_learning_entire}. 
It consists of $T$ deep active learning units, corresponding to $T$
different sensing stages. Each active sensing stage is designed based on an
LSTM cell and a fully connected DNN. 
Specifically, in the $t$-th active sensing stage, the LSTM cell takes the previous cell state vector
$\mathbf{c}_{t-1}$, the previous hidden state vector $\mathbf{s}_{t-1}$, and
the current measurement ${y}_t$ as input, and outputs the next cell state
vector $\mathbf{c}_{t}$ and hidden state vector $\mathbf{s}_{t}$. 
The LSTM cell is capable of automatically summarizing the previous
observations into state vectors. At each stage, we use the fully connected DNN to map
the hidden state vector $\mathbf{s}_{t}$ to the sensing vector $\mathbf{w}_t$.
After receiving the last pilot symbol $y_T$, the LSTM cell updates its cell
state to $\mathbf{c}_T$, which is then mapped to the desired parameter
$\mathbf{v}$ using another DNN. This active sensing framework is trained
end-to-end to maximize the objective function in
\eqref{eq:problem_formulation_unsup}, with neural network weights tied together
across the sensing stages.  Such an end-to-end training approach enables the
learning of an active sensing policy that accounts for the ultimate design or
estimation objective after the $T$ sensing stages.

\subsection{Numerical Results}


\begin{figure}[t]
      \centering
      \includegraphics[width=9.3cm]{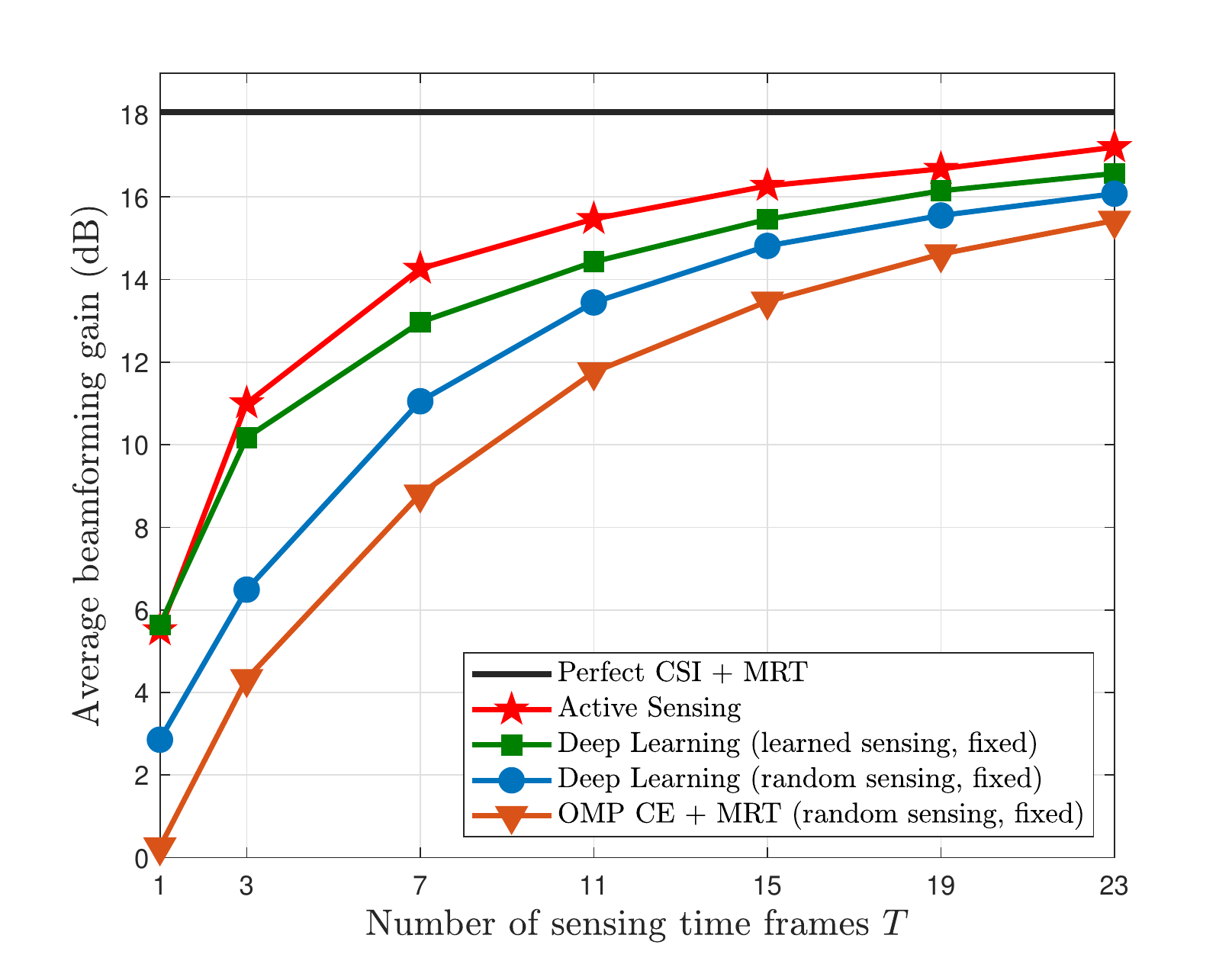} 
      \caption{Average beamforming gain in dB versus the number of sensing stages $T$ for different methods after beam alignment in a TDD mmWave system with
$M_r = 64$, $\operatorname{SNR} = 0$dB, $L_p=3$, and $\phi_1,\phi_2,\phi_3 \in [-60^{\circ},60^{\circ}]$.  \cite{9724252}  }
      \label{fig:sim_multi_AoA}
\end{figure}


To illustrate the performance of the active sensing framework, we now present
the simulation results\footnote{The code for this simulation is available at \url{https://github.com/foadsohrabi/DL-ActiveSensing}}
in \cite{9724252} for a downlink beamforming gain maximization
problem in a setting with $M=64$, $L_{\rm p}=3$, and $\text{SNR}=0$dB. The AoAs are randomly generated from the range $[-60^\circ,60^\circ] $. We compare the proposed active sensing method with the channel estimation based approach as well as a design using a DNN to map the received pilots to the beamforming vector, but the sensing beamformers are fixed, either at random or learned from the statistics of the channel. In Fig.~\ref{fig:sim_multi_AoA}, we see that the deep learning methods outperform the channel estimation based method with OMP. This shows the benefit of bypassing channel estimation. The active sensing method achieves better performance than deep learning with fixed sensing vectors. This shows the benefits of adaptive sensing and the ability of the LSTM network to optimize the sensing vectors.

\begin{figure}[t]
      \centering
      \includegraphics[width=0.226\textwidth]{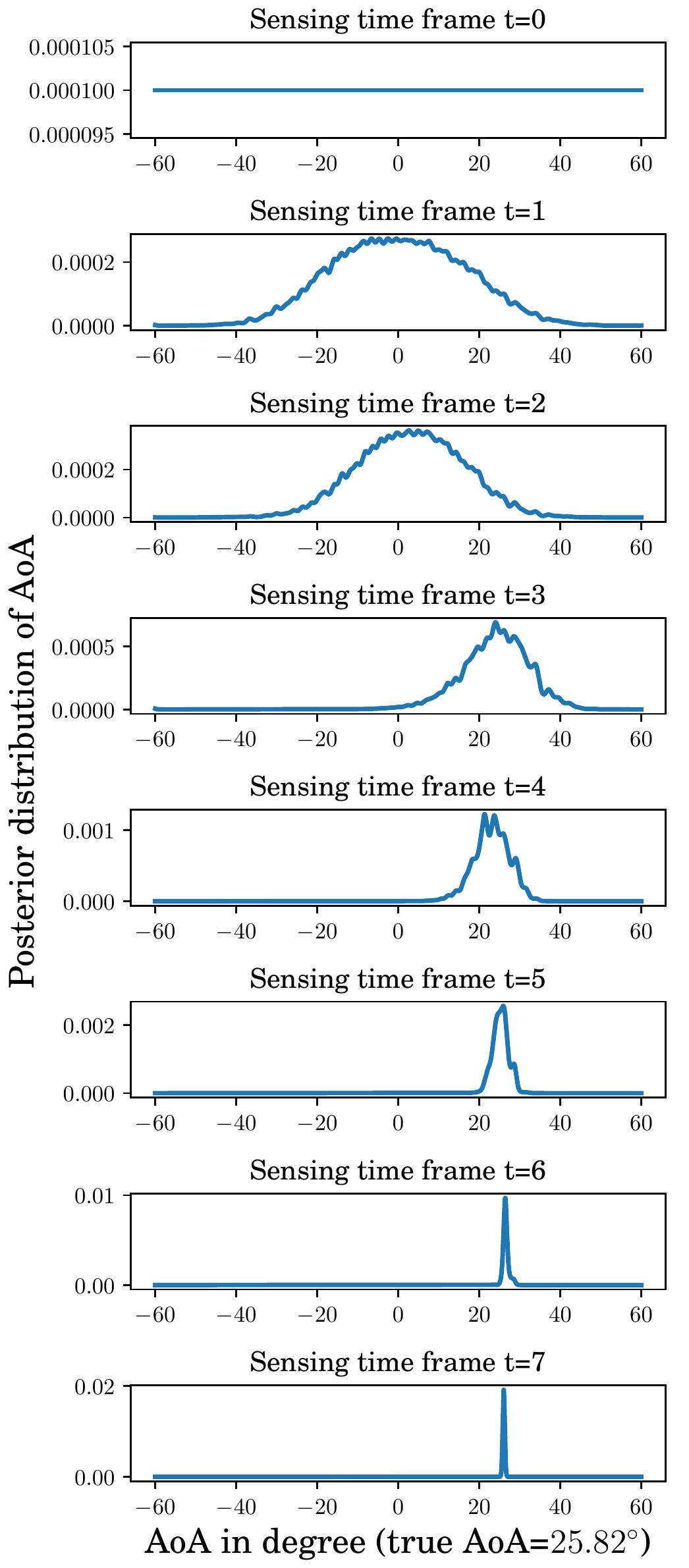} 
      \hspace{0.1cm}
      \includegraphics[width=0.242\textwidth]{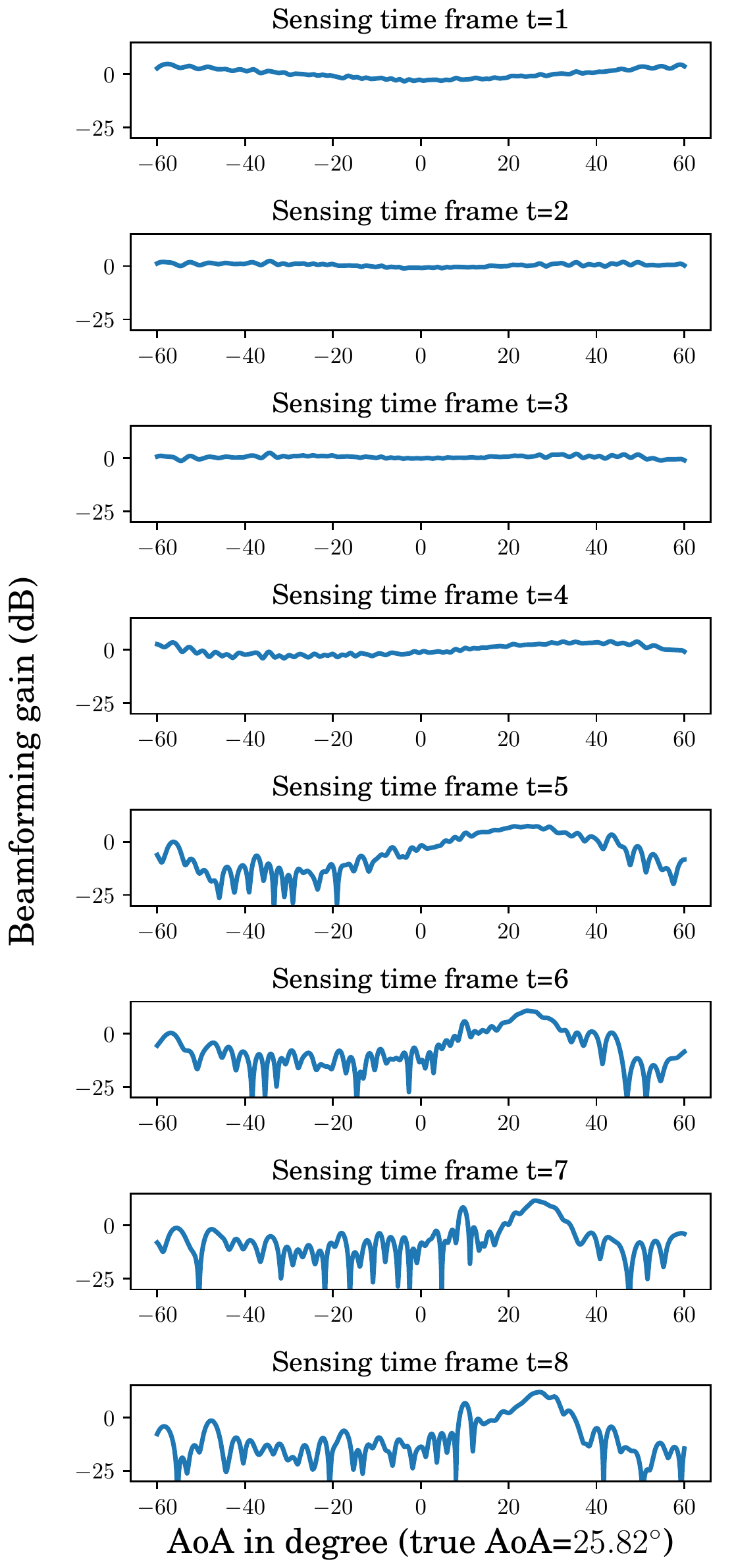} 
      \caption{Posterior distributions of the AoA (left) and the beamforming patterns of the sensing vectors (right) learned from the proposed active sensing framework over $8$ stages in a mmWave alignment problem for a single-path channel where $\operatorname{SNR}=0$dB, $M_r = 64$, and $T = 12$.  \cite{9724252}  }
\label{fig:pos_lstm_tau12}
\end{figure}

To see where the performance gain comes from, we examine the output of the LSTM
framework for an AoA estimation problem in a single-path channel, and plot the
posterior distribution of the AoA at each stage $t$ as well as the array
response of the sensing vectors designed by the LSTM and the DNN. 
As can be seen from Fig.~\ref{fig:pos_lstm_tau12}, the posterior distribution
gradually converges to a distribution concentrated at the true AoA
$\phi=25.82^\circ$.  In the meanwhile, the array response of the sensing
vectors designed by the active sensing framework is relatively flat
across the angles at the beginning, indicating that it is exploring all
directions in searching for the AoA, but gradually narrows down to around the
direction of the true AoA as the sensing operation progresses. This shows that
the active sensing framework indeed learns a meaningful sensing
strategy.  It is capable of quickly converging to the true AoA. It is
remarkable that although finding the truly optimal sensing vectors is extremely difficult 
to do computationally, the LSTM framework is able to learn an excellent sensing
strategy based on training over millions of channel instances.

\section{Standardization Impact}

While the experimental results reported in this article are still generated using widely accepted wireless channel propagation models so the proposed framework should be regarded as proof-of-concept rather than as field tested, the wireless communications standardization bodies have recognized the potential for using machine learning
techniques in future cellular networks and have taken steps toward standardizing the communication protocols between the BS and the users in order to enable learning-based system-level optimization. 

Specifically, the 3rd Generation Partnership Project (3GPP) has recognized channel estimation and feedback, mmWave beam management, and positioning as the three initial areas where machine learning can have a significant impact \cite{3GPP_RP-213599}. 
One of the target scenarios related to the CSI feedback enhancement that 3GPP aims to study is CSI compression and feedback for FDD massive MIMO systems where a wireless device has already obtained the entire high-dimensional channel matrix and it needs to compress and feedback this CSI to the BS. Such a CSI acquisition process can be modelled by an auto-encoder consisting of a DNN encoder at the device and a DNN decoder at the BS. In particular, the DNN encoder first maps the high-dimensional channel to a low-dimensional quantized signal, then the compressed signal is sent to the BS via the uplink feedback channel, and finally, the BS reconstructs the channel by using the DNN decoder. 
The goal of DNNs here is to capture the spatial-domain and frequency-domain correlations in the channel matrix, so convolutional neural networks (CNNs) are an excellent candidate as an autoencoder. Preliminary results reported by the different companies suggest that machine learning can outperform the existing 5G codebook-based CSI compression methods, e.g., \cite{3GPP_R1-2204238}.
This use-case is closely related to the CSI estimation and feedback problem studied in Section \ref{sec:CSI_feedback}.


The second use-case is about beam management procedure (e.g., alignment) to find the best transmit-receive beam pair. The conventional practical beam management is based on exhaustive beam sweeping. While such linear beam search strategies lead to excellent performance, they suffer from significant time delay and power consumption issues.  To address these concerns, sparse beam sweeping has been introduced in 3GPP \cite{3GPP_R1-2203142} in which a beam pair is selected by employing multiple-stage beam narrowing strategies. However, the existing algorithms developed for sparse beam sweeping are quite suboptimal, especially in higher frequency bands (FR2) and with high-mobility users. Data-driven methods, on the other hand, can utilize the training data sets to construct a mapping from sparse beam measurements to the best beam pair. This use-case is closely related to the initial beam alignment problem addressed in Section \ref{sec:active_sensing} using a deep active sensing approach. The use-case can actually be thought of as a non-active version of the problem. 

Accurate positioning is a crucial component in several 5G industrial internet of things (IoT) applications such as smart factories and is another promising area for data-driven designs. The traditional model-based positioning relies on explicit mappings from timing/angle measurements to the position of the user.
These mappings are effective when there are multiple line-of-sight (LoS) paths between the target user and different reception points of the BS. But, practical industrial applications usually have to deal with non-line-of-sight (NLoS) conditions. 
In these scenarios, the traditional model-based approach is not always feasible. Learning-based methods are promising solutions for these difficult positioning tasks since they can easily learn a good mapping from the radio measurements to the position by discerning patterns in the available training data sets. Preliminary results already show significant positioning accuracy enhancement over the conventional methods, e.g., \cite{3GPP_R1-2203901}.
While this article has not addressed the localization problem specifically, the techniques presented are quite applicable to localization \cite{david_GC}.

\section{Conclusion}
In conclusion, the modern machine learning approach is opening new opportunities in the optimization of physical-layer wireless communication systems. It challenges the conventional wisdom of always first modelling the channel, then optimizing wireless system design given the estimated channel. This article shows that much can be gained by bypassing explicit channel modelling, by designing the overall system in an end-to-end manner, and by formulating and solving optimization problems in a data-driven fashion. 

\bibliographystyle{IEEEtran}
\bibliography{IEEEabrv,ref}

\end{document}